\documentclass[aps,prb,onecolumn,showpacs,superscriptaddress,groupedaddress]{revtex4}
\usepackage{epsfig} \usepackage{graphics} \usepackage{subfigure}
\usepackage{amssymb,amsmath}

\begin{document}

\def\e{\begin{equation}}
\def\f{\end{equation}}
\def\_#1{{\bf #1}}
\def\o{\omega}
\def\.{\cdot}
\def\x{\times}
\def\E{\varepsilon}
\def\va{\varepsilon}
\def\M{\mu}
\newcommand{\ds}{\displaystyle}
\def\l#1{\label{eq:#1}}
\def\r#1{(\ref{eq:#1})}
\def\-#1{{\overline #1}}

\def\##1{{\bf#1\mit}}
\def\=#1{\overline{\overline{#1}}}
\def\_#1{{\bf #1}}
\def\^#1{\hat{#1}}
\def\xx{{\ss{\x\atop\x}}}
\def\no{\noindent}
\def\d{\nabla}
\def\D{\partial}
\def\Op{\Omega_{\rm p}}
\def\Om{\Omega_{\rm m}}
\def\Lp{\hat{L}_{\rm p}}
\def\Lm{\hat{L}_{\rm m}}
\def\ap{\alpha_{\rm p}}
\def\am{\alpha_{\rm m}}
\def\b{\delta}
\def\ds{\displaystyle}
\def\ss{\scriptstyle}

\newcommand\vect[1]{\left(\begin{array}{c}#1\end{array}\right)}
\newcommand\matr[1]{\left(\begin{array}{cc}#1\end{array}\right)}

\renewcommand\Re{{\rm Re}}
\renewcommand\Im{{\rm Im}}

\title{Equivalent circuit model of radiative heat transfer}

\author{Stanislav I. Maslovski}

\affiliation{Departamento de Engenharia Electrot\'{e}cnica\\
Instituto de Telecomunica\c{c}\~{o}es, Universidade de Coimbra\\
P\'{o}lo II, 3030-290 Coimbra, Portugal
}

\author{Constantin R. Simovski}
\author{Sergei A. Tretyakov}

\affiliation{Aalto University, School of Electrical Engineering\\
SMARAD Center of Excellence, P.O. Box 13000, 00076 Aalto, Finland
}

\date{\today}

\begin{abstract}
  Here, we develop a theory of radiative heat transfer based on an
  equivalent electrical network representation for the hot material
  slabs in an arbitrary multilayered environment with arbitrary
  distribution of temperatures and electromagnetic properties among
  the layers. Our approach is fully equivalent to the known theories
  operating with the fluctuating current density, while being
  significantly simpler in analysis and applications. A practical
  example of the near-infrared heat transfer through the micron gap filled
  with an indefinite metamaterial is considered using the suggested method. The giant
  enhancement of the transferred heat compared to the case of the empty gap is shown.
\end{abstract}


\maketitle

\section{Introduction}

There are two most important heat exchange processes known: {\em
  thermal conductivity}, associated with collective oscillations of
atoms (i.e., phonons) or electrons in solids including metals, or
with convection in fluids or gases, and {\em radiative heat
transfer} which is associated to the electromagnetic radiation
produced by thermally agitated atoms, e.g., the black body
radiation. In this work we concentrate on the latter process which
can be dominant when the bodies that exchange heat are separated
by gaps (empty or filled with a heterogeneous material that weakly
conducts the heat).

As is known, thermal radiation results from fluctuations of
charge and current density in matter. This phenomenon is governed
by the fluctuation-dissipation theorem~\cite{FDT} (FDT) that relates the
mean-square fluctuations of a physical quantity to the dissipation
associated with the dynamics of the same quantity. Because in
dielectrics the loss is represented by the imaginary part of the
permittivity, which may be in turn related to the effective
conductivity of the material, the FDT requires the volumetric
current density within a dielectric or conducting body to
fluctuate. The FDT constitutes the basis of the present day
radiative heat transfer theories which deal with the fluctuating
currents and treat them as the principal source of thermal
radiation. Such a picture places the radiative heat transfer
calculations in the framework of classical electromagnetic theory
based on the macroscopic Maxwell equations.

Indeed, the well-known theory of radiative heat transfer through
narrow vacuum gaps by Polder and van Hove\cite{Polder} belongs to
this class. Similar techniques have been recently developed for the
cases when radiative heat transfer through micron and even submicron
gaps is assisted with nanostructured metamaterials (see, e.g., in
Refs.~\onlinecite{6,new,arxivTPV}). In these approaches the
fluctuating current density in the two neighboring media is first
obtained from the FDT, following the methodology introduced by
Rytov.\cite{Rytov3} Next, the electromagnetic field produced by the
fluctuating currents is calculated, after which the mean value of the
power flow (Poynting vector) across the gap (filled with the
metamaterial) is found either with the multiple reflection
method,\cite{6,new} or with a more general transfer matrix
approach.\cite{arxivTPV}

Historically, however, thermal agitation of fluctuating currents was
first discovered by Johnson\cite{Johnson} in {\em electric circuits}
and the theory of such thermal fluctuations was developed by
Nyquist\cite{Nyquist} using ideas which were natural for an engineer
dealing with networks of lumped elements and transmission
lines. Although this theory was a precursor to the FDT, it is still in
wide use in the theory of thermal noise at radio and microwave
frequencies. The famous Nyquist's result states that, in any linear
passive two-pole (i.e., a single port device with an input represented
by two electric contacts) operating at a temperature $T$ the electric
thermal fluctuations (in other words, the thermal noise) concentrated
within a narrow frequency interval $\Delta\nu$ can be equivalently
represented by the fluctuating electromotive force (EMF) $e(t)$, with
the mean-square of fluctuations \e\overline{e^2} =
4\Theta(\nu,T)R(\nu)\Delta\nu,\l{Nyquist_intro}\f where
$\Theta(\nu,T)=h\nu/[\exp(h\nu/k_{\rm B}T)-1]$ is Planck's mean energy of a
harmonic oscillator with $h$ and $k_{\rm B}$ being the Planck and Boltzmann
constants, respectively, and $R(\nu)$ is the input resistance (real
part of the input impedance) of the two-pole. The equivalent EMF is
then understood as connected in series with the two-pole, which can be
now considered noiseless.

The beauty of this result is in that {\em no knowledge of the
internal structure of the electric network is required,} and that
all the information is contained within just a single parameter:
the real part of the frequency-dependent input impedance. In the
terminology of FDT, the equivalent EMF in Nyquist's formula has
the role of a generalized force, and the two-pole input impedance
is related to the generalized susceptibility of the system. In
other words, the Nyquist formula can be obtained by a direct
application of FDT to the electric circuit.\cite{Rytov3}  Note
that the fluctuating current appears in this model only as a
reaction to a finite number of {\em lumped} voltage sources.

In contrast, in the thermal transfer theory based on the full-wave
electromagnetic formulation through the fluctuating current density
understood as a {\em volume-distributed} source, the internal
structure of the interacting bodies {\em has to} be considered during
most of the calculations. Thus, a thermal transfer problem appears in
this formulation as a problem with infinite number of degrees of
freedom. The final result, however, happens to be expressed in
quantities that abstract away the internal structure, such as the
reflection coefficients of material half-spaces in the theory of
Polder and van Hove.\cite{Polder} This observation suggests that
introducing the distributed fluctuating current can be avoided in many
cases. For example, in this work we prove that in stratified media the
input impedance concept and the original Nyquist theory can be
generalized and used not only in problems related to electric
networks, but also in full-wave radiative heat transfer problems.
This allows for a significant reduction in complexity of the analysis
and makes the theory readily available for practical calculations.

It has to be mentioned that an expression for the equivalent lumped
EMF of the thermal noise for the case of a lossy material half-space
was first derived by Rytov\cite{Rytov3} using an approach based on the
fluctuating current density. This result constitutes the basis of the
thermal noise theory of aperture antennas.\cite{Pozar,Kraus} Rytov
also worked on equivalent four-pole network representation of a hot
material slab.\cite{Rytov3} Why then these results have not been
widely used in the heat transfer problems? Perhaps, it is because the
concepts of input impedance and the equivalent circuit description for
full-wave electromagnetic problems are largely unknown among theorists
working in the field of radiative heat transfer. In applied
electromagnetics, however, it is well-known that stratified media can
be very efficiently treated within the so-called vector transmission
line theory\cite{Tret} (VTLT) which, in essence, assigns an equivalent
transmission line network to every electromagnetic mode (propagating
or evanescent) in the system. We would like to stress here that the VTLT
is not an approximation: it is a direct consequence of the Maxwell
equations when modal expansion is applied to the electromagnetic field
in layered structures. The VTLT allows also for a systematic treatment
of uniaxial and bi-anisotropic media.

In this work we extend the VTLT in order to include the effect of the
fluctuating current density within the layers. In contrast to a few
numerical discrete-element models currently available from the
literature (e.g., Ref.~\onlinecite{Syms}), the theory that we develop here
is fully analytical. The generalized theory allows us to prove a
complete equivalence between a volumetric multilayered structure and
its circuit theory counterpart, which may be visualized as a chain of
transmission line segments with equivalent fluctuating voltage sources
connected at the ports. When concerned with the radiative heat
transfer between the layers, we show that this equivalent network may
be reduced to just a series connection of a number of voltage sources
representing the fluctuating EMFs and equivalent impedances (each can
be under different temperature), thus, recovering in this way the
famous Nyquist result, generalized here to the full-wave
electromagnetic processes in stratified media.  Therefore, the
calculation of the radiative heat transfer between the layers reduces
in our theory to a number of equivalent circuit theory calculations,
which are relatively simple and very similar to what is typically done
when considering thermal noise in practical electric
networks.\cite{Rytov,Rytov1,Rytov2}

\begin{figure}[!h]
\subfigure[]{\includegraphics[width=0.37\linewidth]{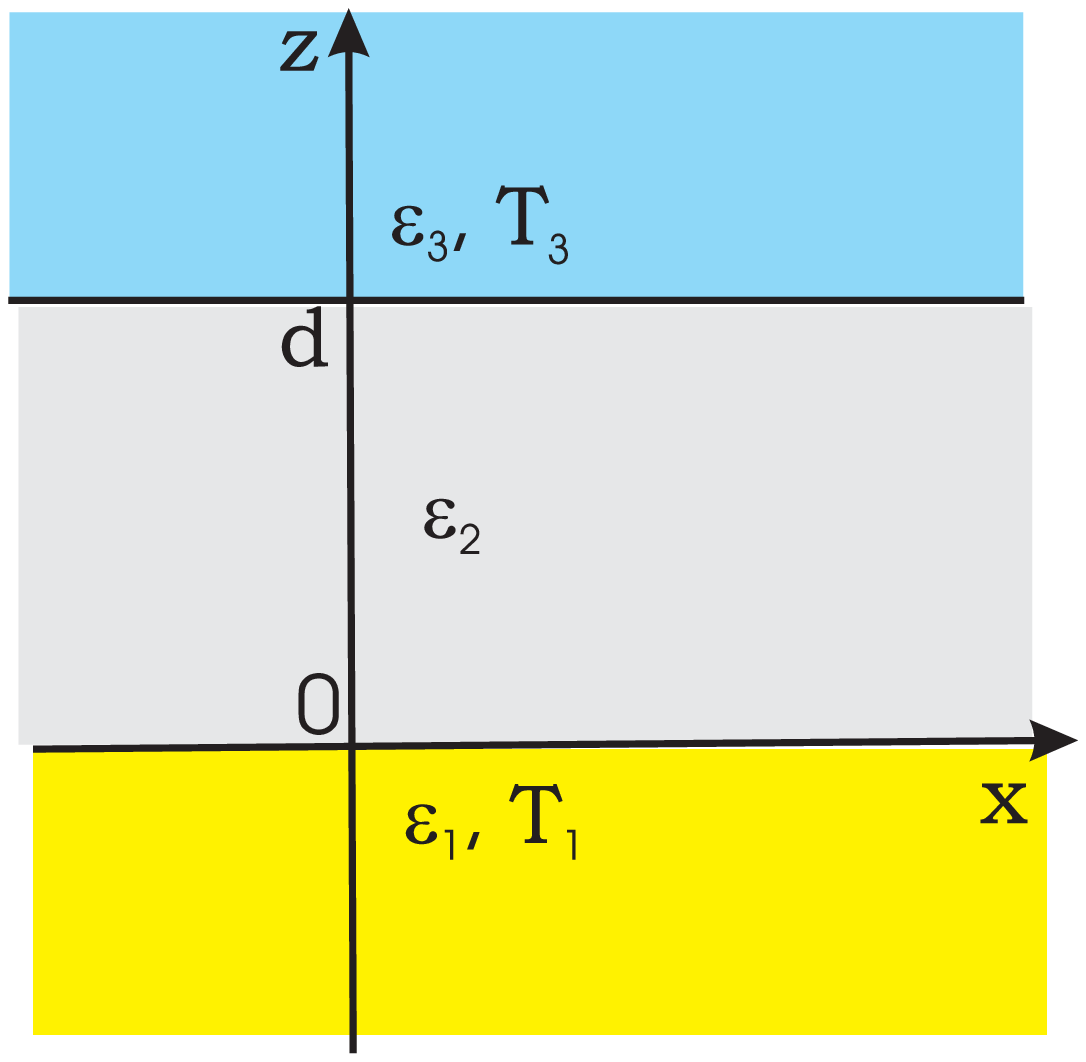}}
\subfigure[]{\includegraphics[width=0.34\linewidth]{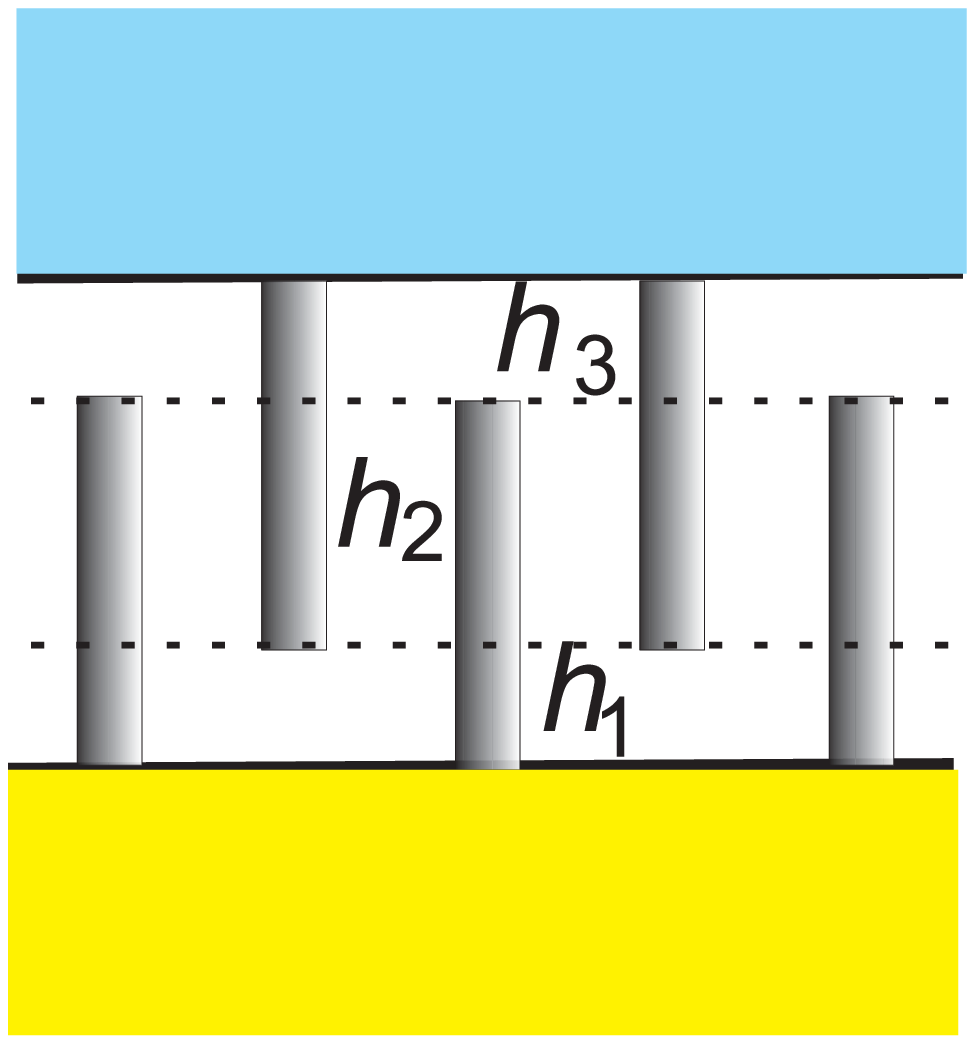}}
\caption{(Color online) (a) -- Illustration to the general problem
formulation. (b) -- A possible implementation of medium 2
suggested in Ref.~\onlinecite{arxivTPV} transforms the gap into a layered structure filled
with so-called hyperbolic metamaterial (see, e.g., in Refs.~\onlinecite{Smith,Narimanov,NarimanovLS,Biehs})
formed by carbon nanotubes or metal nanowires.
Interdigital arrangement of nanotubes (nanowires)
helps keep radiative heat transfer dominating over the thermal
conductance through the gap.}
\label{sh}
\end{figure}

\section{\label{vanHove}Impedance representation of the classical heat transfer formula}

In this section we establish a connection between the classic
Polder--van Hove theory of radiative thermal transfer and its representation in
terms of the input impedances of the material half-spaces.
Formulas by Polder and van Hove for the density of the radiative heat
flux (i.e., power flux of the thermal radiation) across the gap
between two thick dielectric slabs [the geometry is defined in
Fig. \ref{sh}(a); the slabs are approximated by half-spaces] read as
(our notations correspond to Refs.~\onlinecite{5,8,1}): \e
S_{\rm t}=\int\limits_0^{\infty}d\o [\Theta(\o,T_1)-\Theta(\o,T_3)]M,\quad
\Theta(\omega,T_i)=\frac{\hbar\omega}{e^{\hbar\omega \over k_{\rm
      B}T_i}-1}.  \l{Polder}\f Here $\hbar = h/(2\pi)$, and $T_i$ is
the absolute temperature of the $i$-th medium ($i=1,3$; $i=2$
represents the gap). If $T_1>T_3$ the total radiative flux is directed
from medium~1 to medium~3 and can we written as $S_{\rm t}=S_{1\rightarrow
  3}-S_{3\rightarrow1}$
where $S_{1\rightarrow3}$ is the heat flux produced by medium~1 and absorbed in
medium 3 and $S_{3\rightarrow1}$ is the flux produced by medium~3 and absorbed in
medium~1.  $M$ from \r{Polder} is called the radiative heat transfer
function.\cite{1} It depends on the optical properties of all three
media and may be written as the sum $M=M_p+M_e$, where $M_p$ and $M_e$
are contributions of propagating and evanescent waves, respectively
[note that in this paper we assume the time dependence $\exp(+j\o t)$
with $j = \sqrt{-1}$]: \e M_{p}={1\over \pi^2} \int\limits_0^{k_0}
N_p(\o, q) q\, dq, \quad N_p(\o,
q)={(1-|\Gamma_{12}(q,\o)|^2)(1-|\Gamma_{32}(q,\o)|^2)\over
  4|1-e^{-2j\beta d}\Gamma_{12}(q,\o)\Gamma_{32}(q,\o)|^2}, \l{Mp}\f
\e M_{e}={1\over \pi^2} \int\limits_{k_0}^{\infty}N_e(\o, q) q\,
dq,\quad N_e(\o, q)= {{\rm Im}[\Gamma_{12}(q,\o)]{\rm
    Im}[\Gamma_{32}(q,\o)]e^{-2|\beta|d}\over
  |1-e^{-2|\beta|d}\Gamma_{12}(q,\o)\Gamma_{32}(q,\o)|^2}, \l{Me}\f
where $\Gamma_{12}$ and $\Gamma_{32}$ are the reflection coefficients
of a plane wave harmonic having the spatial frequency (transverse wave
number) $q \equiv |\_k_t|$ and being incident from a lossless medium 2
(here, medium 2 is free space) to the surfaces of media 1 and 3,
respectively, $\beta=\sqrt{k_0^2-q^2}$ is the normal component of the
wave vector in medium 2, and $k_0=\o \sqrt{\va_0\mu_0}$ is the wave
number in medium 2. Function $N(\o, q)$ is called the spatial spectrum
of the radiative heat transfer function. This function was studied in
Ref.~\onlinecite{5} for the case when photon tunnelling through the
vacuum gap was enhanced by surface plasmon-polaritons excited at the
gap boundaries. It was shown that the absolute maximum of $N(\o,q)$
achievable at certain values of $\o$ and $q$ equals $1/4$ (we discuss
this limit with more detail in Section~\ref{antenna}).

Let us express reflection coefficients $\Gamma_{12}$ and
$\Gamma_{32}$ in \r{Mp} and \r{Me} through wave impedances
$Z_{1,2,3}$ of the media in regions 1, 2 and 3. The wave impedance
defines the ratio between the transverse electric and magnetic
fields in a plane wave of a given polarization, i.e., in a given
harmonic of the spatial spectrum (see, e.g., in Ref.~\onlinecite{Tret}). In
terms of the wave impedances, \e \Gamma_{i2}= {Z_i-Z_2\over
Z_i+Z_2},\quad i=1,\ 3. \l{imp11}\f For isotropic dielectrics the
wave impedances are given by the following expressions (see, e.g.,
in Ref.~\onlinecite{Tret}) for TM-waves and TE-waves, respectively: \e
Z_i^{\rm TM}=\eta_0{\beta_i\over k_0\va_{i}},\quad Z_i^{\rm
TE}=\eta_0{k_0\over \beta_i}, \l{imp1}\f where $\eta_0$ is the
characteristic impedance of free space
$\eta_0=\sqrt{\mu_0/\va_0}$, and $\beta_i$ denotes the normal
component of the wave vector in the $i$-th medium: $
\beta_i=\sqrt{k_0^2\va_{i}-q^2}$. If medium 2 is free space (as it
is assumed in the classical theory of Ref.~\onlinecite{Polder}) $\va_{i}=1$. For
propagating waves $q<k_0$ and $Z_2$ is real, for evanescent waves
$q>k_0$ and $Z_2$ is imaginary. It is important that the impedance
representation \r{imp11} of the reflection coefficients is general
for all spatial frequencies $q$.

Substituting Eq.~\r{imp11} into Eq.~\r{Mp} we obtain after a
rather simple algebra: \e
N_p=
{16R_2^2R_1R_3\over
  4|(Z_1+Z_2)(Z_2+Z_3)+(Z_1-Z_2)(Z_2-Z_3)e^{-2j\beta_2 d}|^2}.
\l{rewritten1}\f Here and below we denote $R_i\equiv {\rm Re}(Z_i)$
and $X_i\equiv {\rm Im}(Z_i)$. Substituting Eq.~\r{imp11} into Eq.~\r{Me}
we easily deduce: \e N_e= {4X_2^2R_1R_3\,e^{-2|\beta_2|d}\over
  |(Z_1+Z_2)(Z_2+Z_3)+(Z_1-Z_2)(Z_2-Z_3)e^{-2|\beta_2|d}|^2}.
\l{rewritten2}\f In this case $\beta_2$ has imaginary value and it is
taken into account that $Z_2=jX_2$. Since $X_2=0$ for propagating
waves and $R_2=0$ for evanescent waves, Eqs.~\r{rewritten1}
and~\r{rewritten2} can be unified into an expression suitable for both
regions $q<k_0$ and $q>k_0$: \e N(\o,q)= {4|Z_2|^2R_1R_3
  |e^{-j\beta_2 d}|^2\over
  |(Z_1+Z_2)(Z_2+Z_3)+(Z_1-Z_2)(Z_2-Z_3)e^{-2j\beta_2 d}|^2}.
\l{rewritten}\f

The heat flux density originated from the medium 1 and
absorbed in the medium 3 can now be represented as integral over
the complete spatial spectrum (both propagating and evanescent):
\e S_{1\rightarrow3}={1\over \pi^2}
\int\limits_0^{\infty}d\o\int\limits_0^{\infty}qdq\, \Theta(\o,
T_1) N(\o,q). \l{Mep}\f

As we show in the next section, Eq.~\r{rewritten} for the spatial
spectrum of the heat transfer function $N(\o,q)$ can be derived from the
VTLT in a way that allows for a straightforward generalization of the
results of Polder and van Hove\cite{Polder} to the case of stratified
uniaxial magneto-dielectric media {\em without} a need to introduce the
distributed fluctuating currents.

\section{\label{section_circuit}Radiative heat transfer resulting from an equivalent circuit approach}

The possibility to express the radiative heat transfer through the
wave impedances of the material layers provides us with an
evidence that an equivalent circuit model can be formulated for
this problem. Such a model is derived rigorously in Appendix
directly from the Maxwell equations, which results in the VTLT
generalized to include the effect of thermal fluctuations.
Throughout this section, however, we will use simple physical
reasoning when possible, in order to keep the derivations easy to
grasp.

As in the previous section, we decompose the fluctuating
electromagnetic field into plane waves characterized with a
certain polarization state, angular frequency $\o$, and the
transverse wave vector $\_k_t$. For generality, let us assume that
all materials taking part in the heat transfer are optically
uniaxial magneto-dielectric media. It is known that in uniaxial
magneto-dielectrics the independent polarization states correspond
to the TE and TM plane waves. The same holds for a multilayered
structure composed of uniaxial magneto-dielectric layers (some of
them can be isotropic or even vacuum gaps) under the condition
that the anisotropy axes of all layers are aligned. In such a
structure the two polarizations are completely independent and can
be considered separately. In the following we assume that the
anisotropy axis of the layers coincides with the $z$-axis which is
perpendicular to the layers.

In Appendix we prove that, in a given layer of the considered
multilayered structure ($i$-th layer) being under the temperature
$T_i$ the transverse components of the time-harmonic fluctuating
electric and magnetic fields at the layer interfaces (labeled here
with subscripts 1 and 2) are related as follows \e \l{s_zparam} \matr{
  \=Z^i_{11} &\=Z^i_{12}\\
  \=Z^i_{21} &\=Z^i_{22}} \.  \vect{
  \_n_1\x{{\_H}_1^i}_t\\\_n_2\x{{\_H}_2^i}_t} -
\vect{{{\_E}_1^i}_t\\{{\_E}_2^i}_t}= {1\over \sqrt{A_0}}\vect{
  \_e_1^i\\\_e_2^i} \f where $\=Z^i_{mn} = \=Z^i_{mn}(\o,\_k_t)$ are
the dyadic $Z$-parameters of the chosen layer, and $\_e_{1,2}^i$ are
the vectorial fluctuating EMFs equivalently representing the
thermal-electromagnetic fluctuations within the same layer, and
$\_n_{1,2}$ are the external unit normals at the interfaces of the
layer. The meaninig of the factor $1/\sqrt{A_0}$ is explained further
in the text. Eq.~\r{s_zparam} generalizes the known result of the VTLT
to the case of non-vanishing thermal fluctuations. When the right-hand
side of~\r{s_zparam} vanishes this equation represents the definition
of the impedance matrix of a passive material layer.

The result~\r{s_zparam} is obtained in a dyadic form and, thus, is
applicable to any polarization state of the electromagnetic field
in an anisotropic (not only uniaxial) layer. However, when the
states split into the TE and TM waves it is more convenient to
work with scalar $Z$-parameters ---  elements of the $2\times 2$
impedance matrix --- which are defined separately for each
polarization. For a slab of a uniaxial magneto-dielectric
characterized by the permittivity dyadic $\=\E =
\E_i^\perp\=I_t+\E_i^\parallel\_z_0\_z_0$ and the permeability
dyadic $\=\M = \M_i^\perp\=I_t + \M_i^\parallel\_z_0\_z_0$ (here
we understand these parameters as relative to the vacuum
permittivity $\E_0$ and the permeability $\M_0$, respectively,
with $\=I_t$ being the unity dyadic in the transverse plane), the
$Z$-parameters are (see, e.g., Ref.~\onlinecite{Tret}): \e
Z^i_{12}=Z^i_{21}=j{Z_i^{\rm TE,TM}\over
  \sin (\beta^{\rm TE,TM}_i d_i)},\quad Z^i_{11}=Z^i_{22}=-j{Z_i^{\rm
    TE,TM}\over \tan (\beta^{\rm TE,TM}_i d_i)}, \l{zpar} \f where
$\beta^{\rm TE,TM}_i d_i$ is the electric thickness of the $i$-th
layer, and the wave impedances of a spatial harmonic with wave
vector $\_k=(k_x,k_y,\beta^{\rm TE,TM}_i)$ are \e Z_i^{\rm TE} =
\eta_0 {k_0\M_i^\perp\over\beta^{\rm TE}_i}, \quad Z_i^{\rm TM} =
\eta_0 {\beta^{\rm TM}_i\over k_0\E_i^\perp}, \l{_Z_0} \f where the
propagation constants for the two polarizations are expressed
through $q=\sqrt{k_x^2+k_y^2}$ as (see, e.g.,~\onlinecite{Tret}): \e
\beta^{\rm TE}_i =
\sqrt{{\M_i^\perp(k_0^2\E_i^\perp\M_i^\parallel-q^2)\over\M_i^\parallel}},
\quad \beta^{\rm TM}_i =
\sqrt{{\E_i^\perp(k_0^2\M_i^\perp\E_i^\parallel-q^2)\over\E_i^\parallel}}.
\l{_betas}\f

Thus, when considering plane waves of fixed polarization and fixed
transverse wave number $q \equiv |\_k_t|$ such a slab is described by
a $2\x2$ matrix of scalar $Z$-parameters, much like a four-pole
network in the circuit theory.~\cite{Pozar} To make the analogy
complete, we may introduce the effective ``currents'' flowing into
this four-pole network and relate them to the magnetic fields at the
two interfaces of the slab as \e I_{1,2}^{i, \rm TM} =
\sqrt{A_0}\,{\_k_t\over |\_k_t|}\.(\_n_{1,2}\x{{\_H}_{1,2}^i}_t),
\quad I_{1,2}^{i, \rm TE} = \sqrt{A_0}\,{\_z_0\x\_k_t\over
  |\_k_t|}\.(\_n_{1,2}\x{{\_H}_{1,2}^i}_t), \f and the effective
``voltages'' at the input and the output interfaces of the slab \e
V_{1,2}^{i,\rm TM} = \sqrt{A_0}\,{\_k_t\over
  |\_k_t|}\.{{\_E}_{1,2}^i}_t, \quad V_{1,2}^{i,\rm TE} =
\sqrt{A_0}\,{\_z_0\x\_k_t\over |\_k_t|}\.{{\_E}_{1,2}^i}_t.  \f The
factor $\sqrt{A_0}$ where $A_0$ is the unit area in the transverse
plane ensures that the complex power $V^{i,\rm TE,TM}_{1,2}{I^{i,\rm
    TE,TM}_{1,2}}^*= -A_0\,\_n_{1,2}\.\left({\_E_{1,2}^{i,\rm
      TE,TM}}\x{\_H_{1,2}^{i,\rm TE,TM}}^*\right)$ is trivially
related to the complex Poynting vector of a mode.  Then, for a given
polarization (TE or TM) relation~\r{s_zparam} assumes the form \e
\l{s_zmatrscal} \matr{
  Z^i_{11} & Z^i_{12}\\
  Z^i_{21} & Z^i_{22}} \.  \vect{ I_1^i\\I_2^i} - \vect{V_1^i\\V_2^i}
= \vect{ e_1^i\\e_2^i}.  \f

\begin{figure}
\epsfig{file=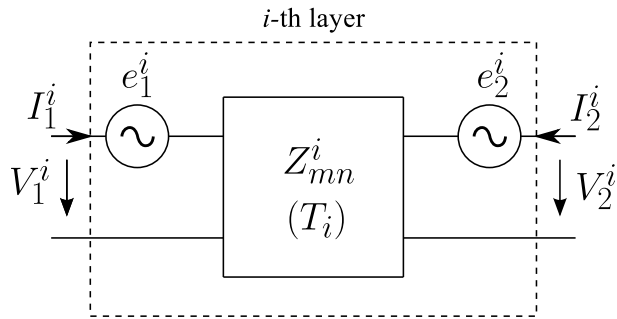,width=0.45\textwidth}
\caption{\label{4pole}Equivalent four-pole network of a material layer under temperature $T=T_i$.}
\end{figure}

The equivalent circuit that corresponds to this relation is shown
in Fig.~\ref{4pole}. The two EMFs at the input and the output of
this circuit represent the effect of thermal fluctuations inside
the chosen layer. The mean square amplitude of these equivalent
sources is derived in Appendix using the approach of the
distributed fluctuating current. However, one can apply the
Nyquist theory directly to the electric circuit in
Fig.~\ref{4pole} and obtain the same result. Namely, by
disconnecting the load from the output of the four-pole network,
i.e., setting $I_2^i = 0$, we eliminate the contribution of
$e_2^i$ and obtain a simple two-pole network with the input
impedance $Z_{\rm in} = V_1^i/I_1^i = Z_{11}^i$. Therefore, the
mean square amplitude\footnote{Throughout this work we use
root-mean square (rms)
  amplitudes for time-harmonic quantities. Thus, if $U(\o)$ is the rms
  amplitude of $u(t)$, then $\overline{u^2(t)} =
  \overline{|U(\o)|^2}$.} of the fluctuating EMF $e_1^i$ within an
angular frequency interval $\Delta\o$ is \e \l{s_e1}
\overline{(e_1^i)^2} = 2\Theta(\o,T_i)\Re(Z_{11}^i){\Delta\o\over \pi}
\f Repeating the same procedure while interchanging the roles of the
input and the output one obtains that \e \l{s_e2}\overline{(e_2^i)^2}
 = 2\Theta(\o,T_i)\Re(Z_{22}^i){\Delta\o\over \pi}
\f It is evident that, in general, the fluctuating sources $e_1^i$
and $e_2^i$ must be partially correlated, because they both
represent the fluctuations within the same layer. Therefore, in
calculations involving expressions which are quadratic in voltage
and (or) current (e.g., power) one may also need the correlation
function of these EMFs: $\overline{(e_1^i e_2^i)}$. As is shown in
Appendix, this correlation can be presented as: \e \l{s_e1e2}
\overline{(e_1^i e_2^i)} =
2\Theta(\o,T_i)\Re(Z_{12}^i){\Delta\o\over\pi}.  \f

The set of relations~\r{s_e1}--\r{s_e1e2} can be also obtained
directly from the FDT. Indeed, if one identifies the charges
$q_{1,2}^i = I_{1,2}^i/(j\o)$ as the state variables of the circuit
depicted in Fig.~\ref{4pole} and the EMFs $e_{1,2}$ as the random
forces associated with the fluctuations, then the FDT demands that for
the fluctuations concentrated within a narrow frequency interval
$\Delta\o$ \e\l{s_FDT} \overline{(e_n^i e_m^i)} = {j\hbar\over
  2}\left[(\alpha^{-1})_{mn}^*-(\alpha^{-1})_{nm}\right]\coth{\hbar\o\over
  2k_B T}\x{\Delta\o\over \pi}, \f where $\alpha_{mn}$ are the
generalized susceptibilities such that $q_m =
\sum_n\alpha_{mn}(\o)e_n$. It is readily seen that
$(\alpha^{-1})_{mn} = j\o Z_{mn}$. Substituting this
into~\r{s_FDT} while taking into account the symmetry properties
of $Z_{mn}$ we obtain~\r{s_e1}--\r{s_e1e2} after dropping the
irrelevant contribution resulting from the quantum zero-point
fluctuations.

\begin{figure}
\epsfig{file=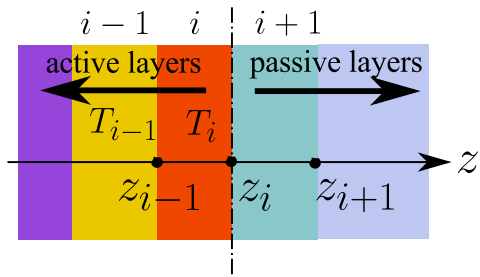,width=0.35\textwidth}
\caption{\label{multilayers} (Color online) An illustration to the
  calculation of the radiative thermal flux through a selected
  boundary $z = z_i$ in a multilayered structure.}
\end{figure}

The relations~\r{s_e1}--\r{s_e1e2} together with~\r{s_zmatrscal}
written for both polarizations fully describe the fluctuations within
a material layer. A structure formed by many layers can now be
equivalently represented by a chain connection of many four-pole
networks each representing a layer. Let us now select an arbitrary
boundary between a pair of layers in a multilayered structure and find
the radiative power flux per unit area of this boundary
(Fig.~\ref{multilayers}).

Because the fluctuating EMFs belonging to separate layers are
uncorrelated, we may first consider only the sources which are located
at $z < z_i$, where $z_i$ is the position of the selected boundary. We
number the layers and the boundaries such that the $i$-th layer is
located at $z_{i-1}\le z \le z_i$. Then, the layers in the half-space
$z > z_i$ can be considered as passive (no radiation is comming out of
them). In the circuit theory terms these layers constitute {\em a
  load} for the other, active, part of the structure located at $z <
z_i$, and can be equivalently represented by an input impedance,
which, for a chain connection of four-pole networks, is given by the
following recursive formula \e Z_{\rm in+}^{i+1} = Z_{11}^{i+1} -
{Z_{12}^{i+1}Z_{21}^{i+1}\over Z_{22}^{i+1} + Z_{\rm in+}^{i+2}},
\l{Zinplus}\f where $Z_{\rm in+}^{i+2}$ is the input impedance of all
the layers behind the $(i+1)$-th layer. The recursion is terminated
with the input impedance of the last layer which extends up to $z =
+\infty$ (it can be, for instance, free space behind the structure),
i.e., with the wave impedance of the last layer. Substituting~\r{zpar}
into~\r{Zinplus} for the two main polarizations in uniaxial layers we
obtain \e Z_{\rm in+}^{i+1} = Z_{i+1}{Z_{\rm
    in+}^{i+2}+jZ_{i+1}\tan(\beta_{i+1}d_{i+1}) \over Z_{i+1} +
  jZ_{\rm in+}^{i+2}\tan(\beta_{i+1}d_{i+1})}, \l{ZinTLplus} \f where,
for brevity, $Z_{i+1} \equiv Z_{i+1}^{\rm TE,TM}$ and $\beta_{i+1}
\equiv \beta_{i+1}^{\rm TE,TM}$. Eq.~\r{ZinTLplus} is well-known in
the theory of transmission lines.

On the other hand, we may apply Th\'{e}venin's theorem to the active
layers located at $z < z_i$. Doing this, one first finds the internal
impedance of Th\'{e}venin's equivalent circuit: \e Z_{\rm in-}^{i} =
Z_{22}^{i} - {Z_{12}^{i}Z_{21}^{i}\over Z_{11}^{i} + Z_{\rm
    in-}^{i-1}}, \l{Zin4pole} \f or, after substituting~\r{zpar}, \e Z_{\rm
  in-}^{i} = Z_{i}{Z_{\rm in-}^{i-1}+jZ_{i}\tan(\beta_{i}d_{i}) \over
  Z_{i} + jZ_{\rm in-}^{i-1}\tan(\beta_{i}d_{i})}, \l{ZinTLminus} \f
where $Z_{\rm in-}^{i-1}$ is the internal impedance of the rest of the
layers located at $z < z_{i-1}$, and the recursion terminates at the
layer which extends down to $z = -\infty$. Note that because here we
consider reciprocal structures, the Th\'{e}venin impedance $Z_{\rm
  in-}^{i}$ equals the input impedance of all the layers located at $z
< z_i$ as seen by a wave incident from the half-space $z > z_i$.

Next, the equivalent voltage generator in Th\'{e}venin's theorem
(recall that the EMF of this generator is the same as the output
voltage of the network under the open circuit condition) can be found
recursively as: \e {\cal E}_{\rm g}^{i} = - e_2^{i} + {Z_{21}^{i}\over
  Z_{11}^{i} + Z_{\rm in-}^{i-1}}(e_1^{i} + {\cal E}_{\rm g}^{i-1}),
\f where ${\cal E}_{\rm g}^{i-1}$ is the equivalent EMF of all the
sources located at $z < z_{i-1}$. This EMF is defined at the boundary
$z = z_{i-1}$. Taking into account relations~\r{s_e1}--\r{s_e1e2} and
the fact that the EMFs corresponding to distinct layers are not
correlated, the mean square amplitude of fluctuations of ${\cal
  E}_{\rm g}^i$ can be expressed after some algebra as
\begin{multline}
\overline{({\cal E}_{\rm g}^i)^2}
=\overline{(e_2^i)^2} + \left|{Z_{21}^{i}\over Z_{11}^{i} + Z_{\rm
      in-}^{i-1}}\right|^2\overline{(e_1^i)^2} -
2\Re\left({Z_{21}^{i}\over Z_{11}^{i} + Z_{\rm in-}^{i-1}}\right)\overline{(e_1^i e_2^i)} +
\left|{Z_{21}^{i}\over Z_{11}^{i} + Z_{\rm in-}^{i-1}}\right|^2\overline{({\cal E}_{\rm g}^{i-1})^2}\\
= 2R^i_{\rm th}\Theta(\o,T_i){\Delta\o\over \pi} + F^i\,\overline{({\cal E}_{\rm g}^{i-1})^2},
\l{Egrec}
\end{multline}
where
\e
F^i = \left|{Z_{21}^{i}\over Z_{11}^{i} + Z_{\rm in-}^{i-1}}\right|^2
={|Z_i|^2\over|Z_i\cos(\beta_id_i)+jZ_{\rm in-}^{i-1}\sin(\beta_id_i)|^2},
\l{sm_noisefact}
\f
and
\e
R^i_{\rm th} = \Re(Z_{\rm in-}^i) - F^i\Re(Z_{\rm
  in-}^{i-1}).\l{defRth}
\f
This important result shows that the effect of thermal
fluctuations within the $i$-th material layer under the temperature
$T_i$ is fully equivalent to the effect of fluctuations in a
resistance $R^i_{\rm th}$ placed under the same temperature.

In other words, Eq.~\r{defRth} manifests that the input resistance
$\Re(Z_{\rm in-}^i)$ of a stack of layers can be split into two
addends: $\Re(Z_{\rm in-}^i)= R^i_{\rm th} + F^i\Re(Z_{\rm
  in-}^{i-1})$. When considered together, these addends represent the
total loss in the stack. However, when the thermal fluctuations are of
concern, Eq.~\r{defRth} allows us to {\em separate explicitely} a part
of the input resistance that appears in the Nyquist formula as being
under physical temperature of the $i$-th layer. Thus, the noise {\em
  produced} within the $i$-th layer is associated with $R^i_{\rm
  th}$. The other addend, $F^i\Re(Z_{\rm in-}^{i-1})$, is due to the
loss in the layers located below the $i$-th layer, and the thermal
noise associated with it is understood as the noise {\em received} by
the $i$-th layer from the background.

Similar concepts exist, for example, in the antenna theory where the
thermal noise of an antenna is represented as a sum of the noise
generated locally by the ohmic loss in the antenna (analogous to
$R^i_{\rm th}$) and the noise received from the environment. The first
addend in this case is proportional to the antenna loss resistance
(which vanishes for an antenna made of a perfect conductor) and the
second term is proportional to the radiation resistance of the
antenna.

From Eq.~\r{defRth}, $\Re(Z_{\rm in-}^{i-1}) = R_{\rm th}^{i-1} +
F^{i-1}\Re(Z_{\rm in-}^{i-2})$, therefore, we may as well write \e
R^i_{\rm th} = \Re(Z_{\rm in-}^i) - F^i R^{i-1}_{\rm th} -
F^iF^{i-1}R^{i-2}_{\rm th} - \cdots, \f where the series terminates at
the layer (with the index $i-M$) that extends to $z = -\infty$, for
which $R_{\rm th}^{i-M} = \Re\left(Z_{\rm in-}^{i-M}\right) =
\Re\left(Z^{\rm TE,TM}_{i-M}\right)$. We may analogously expand the
last addend in~\r{Egrec} which corresponds to the effect of
fluctuations in the layers located at $z < z_{i-1}$. Doing so we
obtain
\begin{multline}
\overline{({\cal E}_{\rm g}^i)^2} = 2R^i_{\rm th}\Theta(\o,T_i){\Delta\o\over \pi} +
2F^iR^{i-1}_{\rm th}\Theta(\o,T_{i-1}){\Delta\o\over \pi} +
2F^iF^{i-1}R^{i-2}_{\rm th}\Theta(\o,T_{i-2}){\Delta\o\over \pi} + \cdots,
\end{multline}
Thus, we conclude that the effect of
thermal fluctuations in {\em all layers} located at $z < z_i$ is the
same as in a chain of resistors with the values $R_{\rm eff}^i =
R^i_{\rm th}$, $R_{\rm eff}^{i-1}=F^iR^{i-1}_{\rm th}$, $R_{\rm
  eff}^{i-2}=F^iF^{i-1}R^{i-2}_{\rm th}$, etc., kept under the
temperatures $T_i$, $T_{i-1}$, $T_{i-2}$, etc. This result is
analogous to the known formula for the thermal noise in cascaded
amplifiers, in which case the quantities $F^i$ are called the noise
factors.

\begin{figure}
\epsfig{file=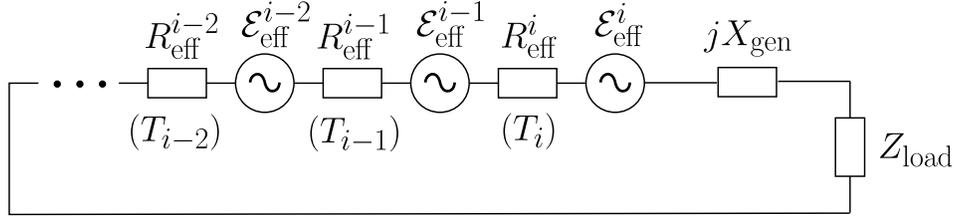,width=0.7\textwidth}
\caption{\label{sm_fig_chain}Th\'{e}venin's equivalent network of
  thermal noise in layered media.}
\end{figure}

The corresponding equivalent circuit is shown in
Fig.~\ref{sm_fig_chain}, in which we split Th\'{e}venin's internal
impedance into a reactive part $X_{\rm gen} \equiv \Im(Z_{\rm in-}^i)$
and a resistive part $\Re(Z_{\rm in-}^i)=\sum\limits_{n}R_{\rm
  eff}^{i-n}$. Respectively, Th\'{e}venin's EMF splits into a series
of uncorrelated fluctuating EMFs: ${\cal E}_{\rm g}^i =
\sum\limits_n{\cal E}_{\rm eff}^{i-n}$, with $\overline{({\cal E}_{\rm
    eff}^{i-n})^2} = (2/\pi)\Theta(\o,T_{i-n})R_{\rm
  eff}^{i-n}\Delta\o$, $n = 0,1,2,\ldots,M,$ representing the effect of
thermal fluctuations in the layers located at $z < z_i$. The rest of
the structure at $z > z_i$ is modeled by an effective load impedance
$Z_{\rm load} \equiv Z_{\rm in+}^{i+1}$.

The radiative heat flux from any layer located at $z < z_i$ into the
half-space $z > z_i$ can now be trivially calculated based on this
equivalent circuit. Namely, the power spectral density associated with
the plane waves with a given transverse wave vector $\_k_t = (k_x,k_y)$
and a given angular frequency $\o$, originated in the slab with the
index $i-n$, can be expressed as \e P^{i-n}_{\o,\_k_t} =
{1\over\Delta\o}{{\overline{({\cal E}_{\rm eff}^ {i-n})^2}} \over
  |Z_{\rm in-}^i + Z_{\rm load}|^2}\Re(Z_{\rm load}) = {2\over\pi}
{\Theta(\o,T_{i-n})R^{i-n}_{\rm eff}\over |Z_{\rm in-}^i + Z_{\rm
    in+}^{i+1}|^2} \Re\!\left(Z_{\rm in+}^{i+1}\right).
\l{dflux_sm}\f Respectively, for the total radiative heat flux
(associated with waves of a selected polarization: TE or TM) into the
half-space $z > z_i$ we have \e S_{z>z_i} =
\sum_{n}\int\limits_0^\infty d\o\iint {dk_x dk_y\over
  (2\pi)^2}\,P^{i-n}_{\o,\_k_t} = {1\over
  2\pi}\sum_{n}\int\limits_0^\infty d\o\int\limits_0^\infty q
dq\,P_{\o,\_k_t}^{i-n}.  \l{flux_sm}\f The flux crossing the the same
boundary in the opposite direction is found by reversing the roles of
the active and passive layers.

\section{Particular case I: Black body radiation}

Let us apply this equivalent circuit theory to calculate the power
radiated by a black body per unit of frequency and surface area. We
assume that a very thick black body lies in the lower half-space $z <
0$ and is under the constant temperature $T$. The upper half-space $z >
0$ is empty. We are interested in the thermal radiation into this
half-space from the black body surface at $z = 0$.

The equivalent circuit for this system is composed of a single
resistance $R_{\rm eff}^{(1)} = \Re\!\left(Z_{\rm
in-}^{(1)}\right)$, where $Z_{\rm in-}^{(1)}$ is the input
impedance of the half-space $z < 0$ occupied by the black body, a
corresponding fluctuating EMF ${\cal E}_{\rm eff}^{(1)}$, a
reactance $X_{\rm gen} = \Im\!\left(Z_{\rm in-}^{(1)}\right)$, and
a load $Z_{\rm load} = Z_{\rm in+}^{(2)} = Z^{\rm TE,TM}_0$, which
is the input impedance of the open half-space at $z > 0$.

By definition, the black body absorbs all incoming radiation
independently of the frequency or the angle of incidence. Thus,
electromagnetically, there are no reflections from such a body which
means that it is perfectly impedance-matched to the free
space. Therefore, $Z_{\rm in-}^{(1)} = Z^{\rm TE,TM}_0$ with \e Z^{\rm
  TE}_0 = {\sqrt{\M_0/\E_0}\over\sqrt{1-q^2/k_0^2}}, \quad Z^{\rm
  TM}_0 = \sqrt{\M_0\over \E_0}\sqrt{1-q^2/k_0^2},\f and, using
Eq.~\r{dflux_sm}, we may write the power spectral density
associated with the radiative heat flux into the open half-space as \e
P_{\o,\_k_t} = {2\Theta(\o,T)\over\pi} {\left[\Re\left(Z^{\rm
        TE,TM}_0\right)\right]^2\over \left|2Z^{\rm TE,TM}_0\right|^2}
= \left\{\begin{array}{ll}
    \Theta(\o,T)/(2\pi), & q \le k_0\\
    0, & q > k_0
\end{array}
\right.  \f The result is the same for both polarizations. From here,
the total power emitted from the black body by both polarizations per
unit of its surface, per unit of frequency is \e {dS\over d\o} =
2\times\iint {dk_x dk_y\over (2\pi)^2}\,P_{\o,\_k_t} =
{\Theta(\o,T)\over 4\pi^3}\iint\limits_{k_x^2+k_y^2<k_0^2}\,dk_x\,dk_y =
{\o^2\over 4\pi^2 c^2}\Theta(\o,T).  \f

The same result can be, of course, obtained from Planck's
expression for spectral radiance of a black body which reads \e
B_\o(T) = {\o^2\over 4\pi^3 c^2}\Theta(\o,T).  \f The spectral
radiance is defined as the power emitted from the black body
surface per unit projected area of the emitting surface, per unit
solid angle, per frequency: \e B_\o(T) = {\Delta P_{\rm rad}\over
A_{\perp}
  \Delta\Omega\,\Delta\o}.  \f Hence, because in spherical coordinates the
projected area $A_{\perp} = A_0\cos\theta$, we find
\e {dS\over d\o} = \int_0^{2\pi}\int_0^{\pi/2} B_\o(T)
\cos\theta\sin\theta\,d\theta\,d\varphi = {\o^2\over 4\pi^2
  c^2}\Theta(\o,T),\f which is the same as the result predicted by the
equivalent circuit model.

\section{\label{genPvH}Particular case II: generalized Polder-van Hove formula}
In this section we derive a generalization of the Polder-van Hove formula
applicable to layered uniaxial magneto-dielectrics, using the
equivalent circuit model developed in
Section~\ref{section_circuit}. The geometry of the structure is the
same as in Fig.~\ref{sh}~(a). We are interested in the radiative
thermal transfer between the media which occupy the half-spaces $z <
0$ and $z > d$ (the media with indices 1 and 3). These half-spaces are
kept under temperatures $T_1$ and $T_3$, respectively. The region
$0<z<d$ (the region 2) may be filled with another uniaxial medium kept
under temperature $T_2$, or may be left empty (which is the case of a
vacuum gap).

In order to find the radiative heat flux from medium 1 into medium
3 we split the structure at the plane $z = d$, and consider the
layers located at $z < d$ as active layers. The half-space $z > d$
plays the role of a load. The equivalent circuit of such a
structure can be represented as in Fig.~\ref{sm_fig_chain}, with a
pair of resistors $R_{\rm eff}^{(1)} = F^{(2)}R_{\rm th}^{(1)} =
F^{(2)}\Re(Z_1)$ and $R_{\rm eff}^{(2)} = \Re\!\left(Z_{\rm
    in-}^{(2)}\right) - R_{\rm eff}^{(1)}$, a pair of the
corresponding fluctuating EMFs ${\cal E}_{\rm eff}^{(1)}$ and ${\cal
  E}_{\rm eff}^{(2)}$, a reactance $X_{\rm gen} = \Im\!\left({Z_{\rm
      in-}^{(2)}}\right)$, and a complex load $Z_{\rm load} = Z_3$.

One may verify that in the vacuum gap case $R_{\rm eff}^{(2)} = 0$,
which is a consequence of the fact that there is no dissipation in the
gap. Evidently, the same conclusion holds when the gap is filled with
a lossless medium. Nevertheless, the following derivation is general
enough to be applicable to both medium-filled or vacuum gaps, with or
without dissipation.

We start with calculating the noise factor $F^{(2)}$. Because $Z_{\rm
  in-}^{(1)} = Z_1$, we obtain from~\r{sm_noisefact}: \e \l{sm_F2} F^{(2)} =
{|Z_2|^2\over|Z_2\cos(\beta_2 d)+jZ_1\sin(\beta_2 d)|^2} =
{|1-\Gamma_{12}|^2|e^{-j\beta_2
    d}|^2\over|1-\Gamma_{12}e^{-2j\beta_2d}|^2}, \f where
$\Gamma_{12}$ is defined in Section~\ref{vanHove}. Thus,
from~\r{dflux_sm}, \e P^{1\rightarrow3}_{\o,\_k_t} =
{2\over\pi} {\Theta(\o,T_1)F^{(2)}\Re(Z_1)\over |Z_{\rm
    in-}^{(2)} + Z_3|^2} \Re(Z_{3}). \l{sm_power}\f
    Next, Th\'{e}venin's internal
impedance $Z_{\rm in-}^{(2)}$ is found from~\r{ZinTLminus}: \e Z_{\rm
  in-}^{(2)} = Z_2{Z_1+jZ_2\tan(\beta_2d) \over Z_2 +
  jZ_1\tan(\beta_2d)} = Z_2{1+\Gamma_{12}e^{-2j\beta_2d}\over
  1-\Gamma_{12}e^{-2j\beta_2d}}, \l{_Zin} \f from which the total impedance of
the network reads \e Z_{\rm in-}^{(2)} + Z_3 = {2Z_2\over
  1-\Gamma_{32}}{1-\Gamma_{12}\Gamma_{32}e^{-2j\beta_2d} \over
  1-\Gamma_{12}e^{-2j\beta_2d}}, \l{sm_totimp} \f where $\Gamma_{32}$
is defined in Section~\ref{vanHove}. Substituting~\r{sm_F2} and~\r{sm_totimp}
into~\r{sm_power} we obtain \e P^{1\rightarrow3}_{\o,\_k_t} =
{\Theta(\o,T_1)\over2\pi}
{|1-\Gamma_{12}|^2|1-\Gamma_{32}|^2|e^{-j\beta_2
    d}|^2\over|1-\Gamma_{12}\Gamma_{32}e^{-2j\beta_2d}|^2} {\Re(Z_1)
  \Re(Z_{3})\over |Z_2|^2}.  \l{sm_power13}\f Respectively, the total radiative heat
flux associated either with TE or TM polarized waves originated
from medium 1 and absorbed in medium 3 is \e S_{1\rightarrow 3}
= {1\over \pi^2}\int\limits_0^\infty d\o\int\limits_0^\infty q dq
\,\Theta(\o,T_1)N(\o,q),\l{sm_StotPvH}\f where \e N(\o,q) =
{|1-\Gamma_{12}|^2|1-\Gamma_{32}|^2|e^{-j\beta_2d}|^2\over
  |1-\Gamma_{12}\Gamma_{32}e^{-2j\beta_2d}|^2}
{\Re(Z_1)\Re(Z_{3})\over 4|Z_2|^2}.
\l{sm_N}\f

In the vacuum gap case, the wave impedance $Z_2$ and the propagation
factor $\beta_2$ are purely real (imaginary) for the
propagating waves (evanescent waves) in the gap. Therefore, because
$Z_i = Z_2(1+\Gamma_{i2})/(1-\Gamma_{i2})$, $i=1,3$, we have
\e
\Re(Z_i) = \left\{
\begin{array}{rl}
|Z_2|{\ds 1-|\Gamma_{i2}|^2\over\ds|1-\Gamma_{i2}|^2}, & \mbox{(propagating waves),} \\[1em]
-|Z_2|{\ds 2\,\Im(\Gamma_{i2})\over\ds|1-\Gamma_{i2}|^2}, &
\mbox{(evanescent waves),}
\end{array}
\right.  \f which results in the Polder-van Hove formulas when
substituted into~\r{sm_StotPvH} and~\r{sm_N}. However, the more
general result represented by Eqs.~\r{sm_StotPvH}--\r{sm_N} holds
for {\em arbitrary} uniaxial magneto-dielectric media filling the
gap. It is easy to verify that Eq.~\r{sm_N} can be as well written
in form~\r{rewritten}.

Moreover, in general, when the gap is filled with a lossy medium and
$T_2\neq 0$ one also has to take into account the radiative heat flux
to medium 3 that is originated in the gap (i.e., in medium 2). The
power spectral density associated with it is, from Eq.~\r{dflux_sm},
\e P^{2\rightarrow3}_{\o,\_k_t} = {2\over\pi}{\Theta(\o,T_2)R_{\rm
    eff}^{(2)}\over |Z_{\rm in-}^{(2)} + Z_3|^2} \Re(Z_{3}) =
{\Theta(\o,T_2)\over2\pi}
{|1-\tilde{\Gamma}_{12}|^2|1-\Gamma_{32}|^2|\over|1-\tilde{\Gamma}_{12}\Gamma_{32}|^2}
{R_{\rm eff}^{(2)}\Re(Z_{3})\over |Z_2|^2}, \l{sm_gapflux} \f where
$\tilde{\Gamma}_{12} = \Gamma_{12}e^{-2j\beta_2d}$ is the reflection
coefficient defined at the plane $z = d$. One may note
that~\r{sm_gapflux} has the same form as~\r{sm_power13} with $d = 0$
and one of the wave impedances replaced by the effective resistance
$R_{\rm eff}^{(2)}$. This is the consequence of the fact that when the
thickness of the middle layer increases, $\tilde{\Gamma}_{12}\rightarrow 0$,
$R_{\rm eff}^{(2)}\rightarrow \Re(Z_2)$, and~\r{sm_gapflux} reduces to
the Polder-van Hove's result for two media in direct contact.

\section{\label{antenna}Antenna theory and circuit theory concepts applied to radiative heat transfer}

In this section, in order to better understand the circuit model
developed in this work, we establish a connection between our model
and the classical theory of noise in receiving antennas. We consider
the radiative heat transfer example from Section~\ref{genPvH} and
employ an analogy between the heat-receiving half-space (medium 3) and
a loaded receiving antenna.  More exactly, in this analogy the unit
area of the interface between the media~2 and~3 is treated as an
aperture antenna that receives the power of thermal radiation from
the half-space $z < d$ and delivers it to medium 3 which is
understood as the antenna load. In what follows, we assume that
medium~2 is lossless, therefore, all the radiative heat delivered to
medium~3 is generated in medium~1.

The equivalent circuit of this problem is the same as the one
discussed in Section~\ref{genPvH} with $R_{\rm eff}^{(2)} = 0$. Thus,
in the antenna analogy there is one noise source with the internal
impedance $Z_{\rm in-}^{(2)} \equiv Z_{\rm A} \equiv R_{\rm A} +
jX_{\rm A}$, where $X_{\rm A} = X_{\rm gen}$ is the antenna reactance
and $R_{\rm A} = F^{(2)}\Re(Z_1)$ is the {\em radiation resistance} of
the antenna. Such analogy signifies that the effective radiation
resistance of an aperture equals the real part of the input impedance
of the half-space seen from the aperture. The antenna load is
represented in this analogy by the impedance $Z_{\rm load} = Z_3$
which is the wave impedance of medium~3.

As is known from the theory of noise in lossless antennas\cite{Pozar}
(an aperture by itself has no loss), the mean-square EMF
$\overline{e^2}$ of the thermal noise of a directive antenna (e.g., a
radio telescope) is given by the Nyquist formula~\r{Nyquist_intro}, in
which one inserts the radiation resistance of the antenna [as
$R(\nu)$] and the effective temperature of the area of the sky to
which the antenna is directed (which is medium~1 under temperature $T
= T_1$ in our analogy). Applying this to the antenna equivalent
circuit, we may write for the noise power at the antenna load: \e
P_{\rm out} = {\overline{e^2}\over|Z_{\rm A} + Z_3|^2 }\Re (Z_{3}) =
{2\over\pi} {\Theta(\o,T_1)R_{\rm A}\over |Z_{\rm A} + Z_3|^2} \Re
(Z_{3}).\l{_p}\f For our example of a lossless medium 2, $R_{\rm
  A}=\Re\left(Z_{\rm in-}^{(2)}\right)$, which brings us to the same
result as the more general cascade-circuit model~\r{sm_power}, i.e.,
$P_{\rm out} = P^{1\rightarrow3}_{\o,\_k_t}$. Let us also note that if
there would be no separating layer, then $Z_{\rm A}$ would be simply
equal to the wave impedance of medium~1.  With the separation layer in
place, antenna model calculation is equivalent to calculation of input
impedance of a transmission-line section loaded with a known impedance
[Eq.~\r{_Zin}].

It is worth noting that the tight connection between the model of the
present paper and the theories of noise in antennas and cascaded
electric networks allows one to better understand optimal conditions
for radiative heat transfer through composite layers and,
consequently, to design these material structures aiming for desired
and optimized performance. In particular, from \r{_p} and \r{sm_power}
we see that the problem of maximizing radiative heat transfer for
given layer temperatures reduces to an equivalent problem of matching
a generator to a load. Let us discuss this issue assuming for
simplicity, that the media~1 and~3 are the same, i.e., $Z_1=Z_3$. At
the first glance, it appears that the optimal heat transfer is ensured
if we simply connect the two equivalent media together (or fill the
gap with the same medium as those~1 and~3). However, this is true only
if the wave impedance is real. For complex $Z_1$ and $Z_3$ (which is
realistic even for propagating modes in view of losses in the media),
the best radiative heat transfer corresponds to the conjugate
impedance matching $Z_1=Z_3^*$ when the negative reactance of one of
two media is compensated by the positive reactance of the other
one. Then, the spatial spectrum of the heat transfer function in
accordance to \r{rewritten} turns to $N=1/4$ whereas for the direct
contact of two equivalent media with wave impedance $Z=R+jX$ we have
$N=R^2/|2Z|^2$. This tells us that the radiative heat transfer through
a properly filled gap can be in principle made larger than that
through the direct contact of two equivalent media.

In Ref.~\onlinecite{arxivTPV} it was proposed to insert a slab of the
so-called indefinite medium between media 1 and 3 [Fig. \ref{sh}~(b)].
Indefinite media (also called hyperbolic metamaterials\cite{Smith})
are uniaxial dielectrics characterized by the permittivity tensor that
has opposite signs of the longitudinal and transverse components. Such
filling allows for enhancement of the heat transfer by increasing the
noise factor $F^{(2)}$. This factor increases compared to the vacuum
gap because indefinite media support propagation of spatial harmonics
with high transverse wavenumbers, which would otherwise be evanescent
in the gap [notice the exponential factor in Eq.~\r{sm_F2}].  For such
waves, $F^{(2)}$ dramatically increases at the spatial frequencies
that correspond to the minima of the denominator of Eq.~\r{sm_F2}.
However, it is not straightforward to ensure proper impedance match in
such structures. In this view, the structure suggested and studied in
Ref.~\onlinecite{arxivTPV} is not fully optimal, although it still
demonstrates that a specifically crafted filling may dramatically
enhance the radiative heat transfer through the gap.

The transmission line analogy, however, suggest an immediate
possibility to circumvent the problem of impedance conjugate
match. The key is to make the wave impedances of all three media
{\itshape real} (at least, approximately) and equal in all
layers. This can be realized using a wire medium in region 2 which
extends inside regions 1 and 3. Indeed, it is known that wire media
support propagating TEM modes with high spatial frequencies
(transverse wave numbers) $q$, including $q>k_0$ and limited only by the
period of the wire array (see, e.g., in Ref.~\onlinecite{AM}). Wires
should be good conductors at the frequency of interest, and the
background media nonconducting. If the wires extend over all three
regions and none of them has a conducting background, then these TEM
modes will exist and have real wave impedance everywhere in the
system, in principle allowing the conjugate match of the load to the
heat source for a wide range of spatial frequencies $q$.

In the present paper we, however, do not present the studies of the
heat transfer optimization keeping these results for our next
publications. Instead, in the next section we report some numerical
results illustrating the applicability of our model for solving
practical problems.

\section{Numerical example: Radiative thermal transfer through a nanostructured layer}

In order to demonstrate the applicability of the developed theory to
real-world problems, we calculate in this section the spectral density
of the radiative heat flux absorbed in medium 3 of the structure
depicted in Fig.~\ref{sh}~(b): \e s_{13}\equiv {dS_{1\rightarrow
    3}\over d\o}= {1\over 2\pi}\int\limits_{0}^{\infty}
P^{1\rightarrow3}_{\o,\_k_t}\, qdq. \l{s13}\f We compare the heat
transferred across the vacuum gap with that transferred across the gap
filled with either normally oriented metal-state single-wall carbon
nanotubes (CNT) or with similarly oriented golden (Au)
nanowires. Calculation for the array of CNT is done in order to
validate the present model using the exact simulations of
Ref.~\onlinecite{arxivTPV}. Calculations for metal nanowires are done
in order to confirm or decline the effect of giant enhancement of
radiative heat transfer in the near infrared (IR) range due to the
presence of nanowires. This effect was predicted but not studied in
Ref.~\onlinecite{arxivTPV}. In accordance with the theory presented
above, all calculations are done for homogenized media.  The
homogenization models for $i$-th medium results in explicit formulas
for $\E_i^\perp$ and $\E_i^\parallel$ (whereas $\M_i^\perp=\M_i^\parallel=1$).

The mid-IR homogenization model for the array of CNT was described and
validated in Ref.~\onlinecite{Nefedov}. The parameters of the array of
CNT correspond to those of Ref.~\onlinecite{arxivTPV} [see also in
Fig.~\ref{sh}~(b)]. The homogenization model for aligned metal
nanowires representing their array as a layer of an indefinite
material was described in Ref.~\onlinecite{Elser}. It is applicable to
the visible and near-IR ranges where it offers a rather high accuracy
for optically dense arrays of rather thin wires. Practically, for the
band $\lambda=$1--2$\ \mu$m one needs the period below 300--600~nm and the
wire diameter of the order of the skin depth in the metal or
smaller.\cite{AM} For Au in this wavelength range it means that the
thickness of the wires should not exceed 20--50~nm.

\begin{figure}[!h]
\subfigure[]{\includegraphics[width=0.45\linewidth]{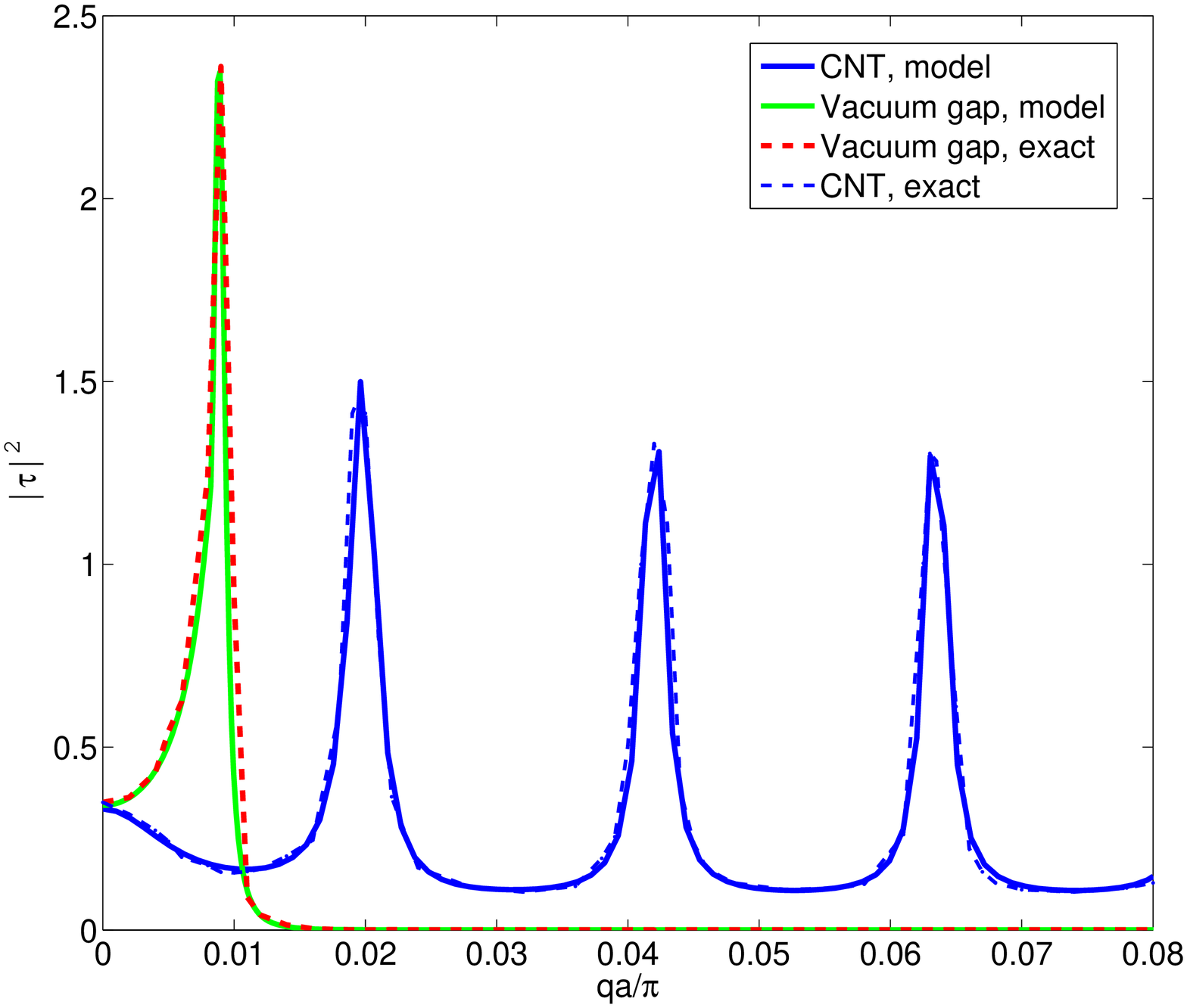}}
\subfigure[]{\includegraphics[width=0.45\linewidth]{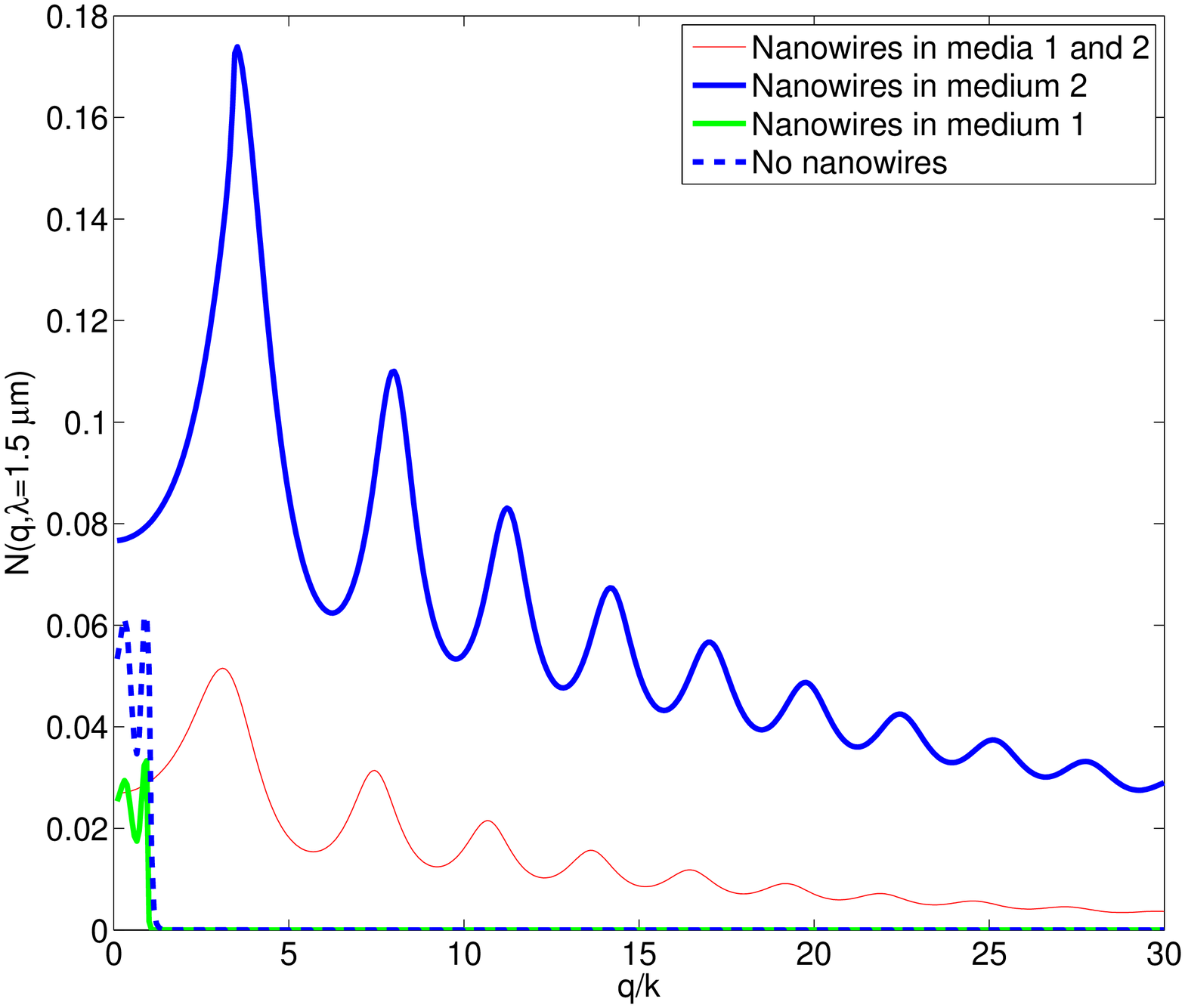}}
\caption{(Color online) (a) -- Coefficient $|\tau|^2=4|Z_1/R_1|^2 N$ at
$\lambda=7.5\ \mu$m versus normalized spatial frequency $qa/\pi$
for the gap filled with free space and for that filled with an
array of CNT. Dashed lines -- exact (beyond homogenization)
simulations from Ref.~\onlinecite{arxivTPV}. Solid lines --
analytical calculations in accordance to the present model. (b) --
Spatial spectrum $N$ of the radiative heat transfer function
versus dimensionless spatial frequency $q/k$ at $\lambda=1.5\ \mu$m. Calculations are
done for four cases -- nanowires are only in medium 2, nanowires
are in both media 1 and 2, nanowires are in medium 1 only and
nanowires are absent.}
 \label{Sim}
 \end{figure}

 In both cases (CNT in the mid-IR range and nanowires in the near IR)
 the presence of the metamaterial enhances the radiative heat
 transfer. This effect results from the conversion of TM-polarized
 evanescent waves into propagating ones in the effective indefinite
 material filling the gap.\cite{arxivTPV,NarimanovLS,Biehs}
 Respectively, in this section we analyze only that part of the
 radiative heat which is transferred by TM-polarized waves.

\begin{figure}[!h]
\subfigure[]{\includegraphics[width=0.45\linewidth]{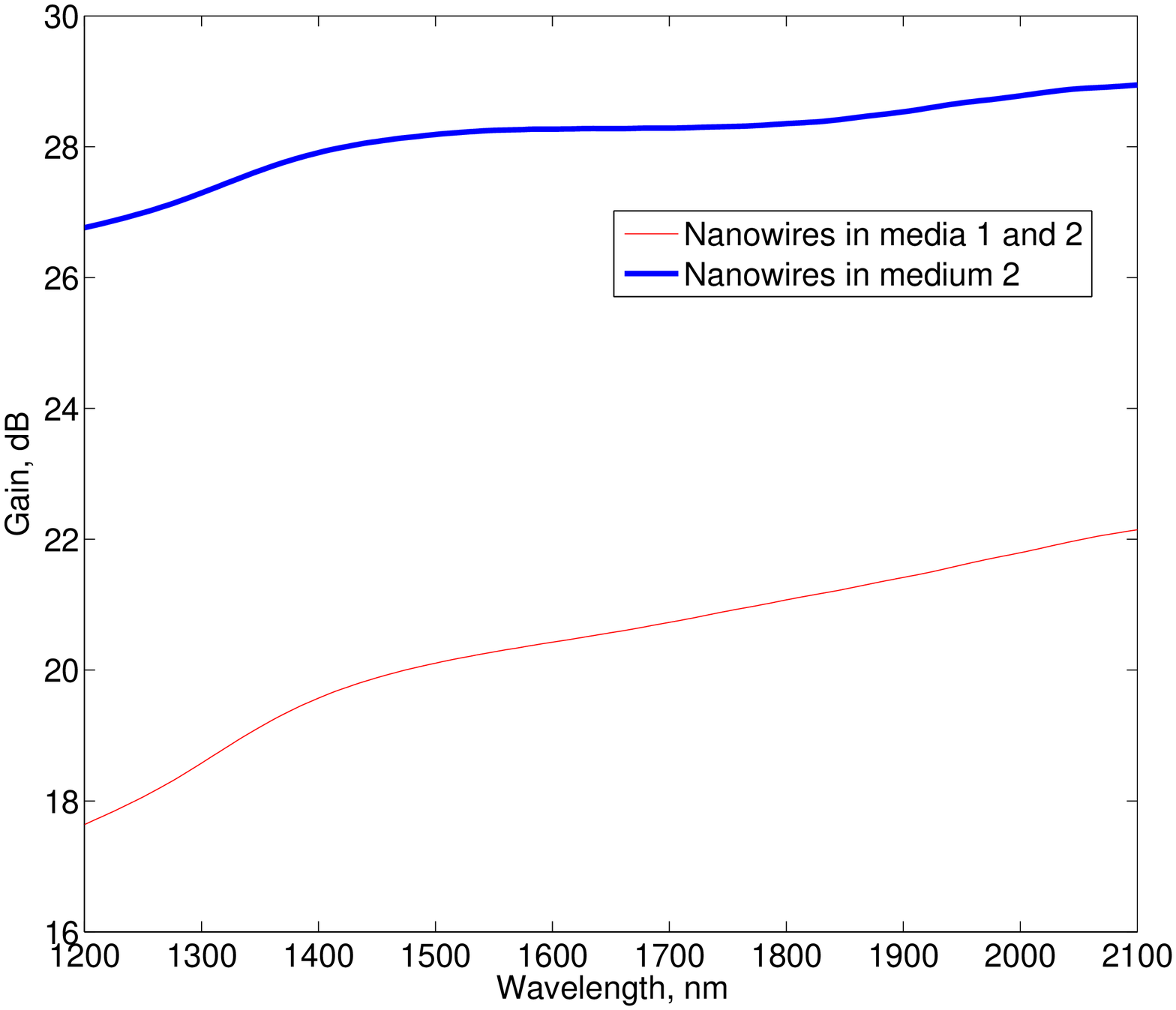}}
\subfigure[]{\includegraphics[width=0.46\linewidth]{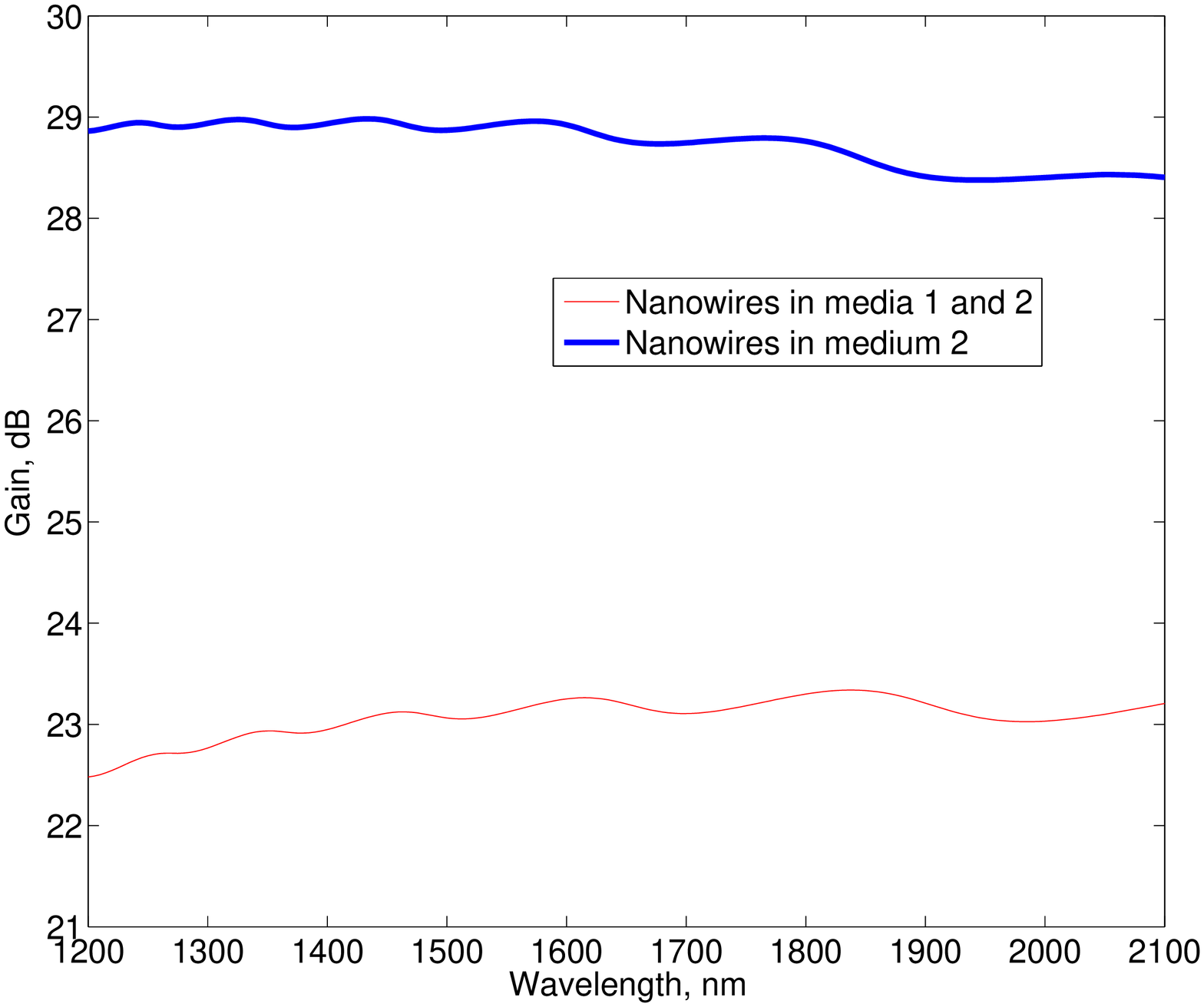}}
\caption{(Color online) The gain in the spectral density of the
transferred radiative heat flux due to the presence of nanowires
(two cases of their arrangement). (a) --   Gap of thickness $d=2\
\mu$m. (b) -- Gap of thickness $d=0.5\ \mu$m.}
 \label{Sim1}
\end{figure}

In this numerical example we neglect the contribution of thermal
sources located in medium 2 (i.e., in CNT and nanowires). The impact
of thermal production and absorption in medium 2 will be evaluated in
our next paper, where we will also consider thermo-photovoltaic
applications of our present model. As is clear from Fig.~\ref{sh}~(b),
medium 2 in the gap is a triple-layered structure, therefore, there
are in total five material layers in the whole structure. However,
since in this example we neglect the thermal processes in the gap, we
may replace the layers in the gap by a single equivalent four-pole
network. Its $Z$-matrix, $Z_{mn}^{(2)}$, is obtained in a standard
manner from its transfer matrix,\cite{Tret} with the latter being a
product of transfer matrices of effectively homogeneous anisotropic
layers with thicknesses $h_1$, $h_2$ and $h_3$. Next, the power
spectral density $P^{1\rightarrow3}_{\o,\_k_t}$ in Eq.~\r{s13} is
calculated from Eq.~\r{sm_power} where the noise factor is $F^{(2)} =
\left|Z_{21}^{(2)}\right|^2/\left|Z_{11}^{(2)} + Z_1\right|^2$
[Eq.~\r{sm_noisefact}], and the internal impedance is $Z_{\rm
  in-}^{(2)} = Z_{22}^{(2)} - Z_{12}^{(2)}Z_{21}^{(2)}/\left(Z_{11}^{(2)} +
Z_1\right)$ [Eq.~\r{Zin4pole}].

In Fig.~\ref{Sim}~(a) we depict the energy transfer coefficient
$|\tau|^2$ introduced in Ref.~\onlinecite{arxivTPV} [Eq.~(12) of
Ref.~\onlinecite{arxivTPV}] which differs from our heat transfer
spatial spectrum $N(\o, q)$ defined by Eq.~\r{Mep}, by the factor
$4|Z_3|^2/R_1R_3$. In the present case media 1 and 3 are equivalent
(heavily doped silicon), i.e., $Z_1=Z_3,\ R_1=R_3$.
All parameters in the calculation illustrated by
Fig.~\ref{Sim} (a) correspond to those from
Ref.~\onlinecite{arxivTPV}.  The value $|\tau|^2$ is calculated at
wavelength $\lambda=7.5\ \mu$m as a function of the normalized spatial frequency
$qa/\pi$, where $a=20$ nm is the period of the CNT array in the
domains $h_1=d/3$ and $h_3=d/3$, where the gap thickness is $d=1\
\mu$m. In the domain $h_2=d/3$ the array period is equal
$a_2=\sqrt{20}$ nm.  Heavily doped silicon supports so-called surface
plasmon-polariton (SPP) waves generated on the surfaces of the Si
half-spaces at $(qa/\pi)=0.01$ in the case when the gap is empty.  The
value $q=0.01\pi/a$ nearly corresponds to $q=1.8k_0$, where $k_0$ is
the free-space wave number.  The manifestation of this SPP is the
maximum of the corresponding curve in Fig.~\ref{Sim}~(a).  Dashed
curves in Fig.~\ref{Sim}~(a) correspond to exact simulations of
Ref.~\onlinecite{arxivTPV} which take into account the microstructure
of the material layer in the gap.
In Fig.~\ref{Sim}~(a) we show only the region of spatial frequencies $(qa/\pi)\le 0.08$ in
which the difference between the exact and homogenized models of the
CNT array is negligibly small (see Ref.~\onlinecite{arxivTPV}).
For both empty gap and gap filled
with CNT the agreement between our model and the exact simulations is
excellent. Local maxima of the transmittance spatial spectrum for the
gap filled with CNT correspond to the thickness resonances of spatial
harmonics [in presence of CNT the whole region $(qa/\pi)\le 0.08$ corresponds to
propagating plane waves, though the inequality $q>k_0$ holds for $(qa/\pi)>5.6\cdot 10^{-3}$].
Fine agreement between our circuit model and exact simulations
pertains at other wavelengths besides the ones mentioned in
Figs.~\ref{Sim}~(a,b), because with our circuit model we also
have reproduced the frequency dependence of the gain in the heat
transfer $G(\o)=s_{13}^{(CNT)}(\o)/s_{13}^{(0)}(\o)$, calculated in
Ref.~\onlinecite{arxivTPV}. Here $s_{13}^{(0)}$ corresponds to the
vacuum gap and $s_{13}^{(CNT)}$ corresponds to the gap filled with
CNT.

In Fig.~\ref{Sim}~(b) we present the dependence $N(q,\lambda^*)$ for
the case when the gap $d=2\ \mu$m is filled with golden nanowires. The
complex permittivity of gold in the range $\lambda=1\dots 2\ \mu$m was
taken from Ref.~\onlinecite{JC}.  Regions 1 and 3 in this case are
filled with doped germanium used in real thermo-photovoltaic systems
whose complex permittivity was taken from
Ref.~\onlinecite{Ge}. Function $N(q,\lambda^*)$ was calculated for
nanowires with volume fraction $p=0.2$ in the domains $h_1$ and $h_3$
and $p=0.4$ in the domain $h_2$ (when $h_1=h_2=h_3$). Here
$\lambda^*=1.5\ \mu$m has been chosen having in mind possible
thermophotovoltaic applications (at this wavelength the doped Ge has
nearly maximal photovoltaic spectral response). The result for $N$
(thick solid curve) was compared with that for the empty gap $d=2\
\mu$m (thin dashed curve) at the same wavelength.

It has to be mentioned that the structure shown in Fig.~\ref{sh}~(b)
with free-standing metal nanowires is an abstraction --- in a feasible
structure nanowires are partially submerged into the host
material. Therefore we have also calculated the function
$N(q,\lambda^*)$ for the case when Au nanowires are semi-infinite and
have the same volume fraction $p=0.3$ in media 1 and 2. This
calculation [thick dashed curve in Fig.~\ref{Sim}~(b)] is done in
order to understand how the extension of nanowires into medium 1
changes the radiative heat transfer. In this case the gap is uniformly
filled, i.e., in the the structure shown in Fig.~\ref{sh}~(b),
$h_2=h_3=0$ and $h_1=d$. In this case the nanowres touch the surface
of medium 3 and the thermal transfer by the direct thermal conductance
may be of significance, in addition to the radiative one. This effect
is not considered in the present paper. Additionally, we have studied
the case when nanowires with $p=0.3$ are located only in medium 1 and
the gap is empty [thin solid curve in Fig.~\ref{Sim}~(b)].

We can see in Fig.~\ref{Sim}~(b) that the integral increase of $N(q)$
with respect to the empty gap is very significant for both cases when
the nanowires fill in the gap. Function $N(q)$ in the case of the
vacuum gap has two local maxima in the region $q<k_0$ resulting from
Fabry-Perot resonances. Because the structure is not fully
impedance-matched these maxima are much smaller than the achievable
limit $N=1/4$ and $N(q)$ vanishes fast at $q>k_0$. Unlike the
situation illustrated by Fig.~\ref{Sim}~(a) the real part of the
complex permittivity of Ge at $\lambda=1.5\ \mu$m is positive and SPP
cannot be excited.  Since the value of $N(q)$ for the empty gap is so
small, the gain granted by Au nanowires to the heat transfer between
two half-spaces of Ge turns out to be larger than that offered by CNT
to the heat transfer between two half-spaces of Si calculated in
Ref.~\onlinecite{arxivTPV}.

The presence of nanowires only in medium 1 turns out to be destructive
for the amplitude of $N$. In this case medium 1 is an indefinite
metamaterial, and the mismatch between medium 1 and free space
increases.  However, if nanowires are present in both media 1 and 2,
noticeable values of $N$ keep for $q>k_0$. The  case when
nanowires are located only in medium 2 is the best one: it corresponds
to the smallest mismatch between the media.

The dependencies shown in Fig.~\ref{Sim}~(b) are typical for every
$\lambda$ in the band of the photovoltaic operation of Ge
($\lambda=1-2\ \mu$m). As a result, due to the presence of nanowires
the heat transfer gain $G=s_{13}^{(NW)}/s_{13}^{(0)}$ is almost
uniform over a wide range of wavelengths. Here $s_{13}^{(NW)}$
corresponds to the case of nanowires in the gap and $s_{13}^{(0)}$
corresponds to the vacuum gap. In Fig.~\ref{Sim1}~(a) we present the
gain in dB, i.e., $10\log_{10} G$ calculated for the case $d=2\
\mu$m. The huge gain keeps for the interval of values $d=0.5\dots 5\
\mu$m. In Fig.~\ref{Sim1}~(b) we show the same gains for the case
$d=0.5\ \mu$m. We can conclude that the presence of Au nanowires in
the micrometer or submicron gap can increase the near-IR energy
transfer across the gap by 3 orders of magnitude. This result confirms
the expectations of Ref.~\onlinecite{arxivTPV}.

\section{Conclusions}

In this work we have formulated an equivalent circuit theory of the
radiative heat transfer in uniaxial stratified magneto-dielectric
media. We have proven that the effect of thermal-electromagnetic
fluctuations in such structures can be fully determined without an
explicit knowledge of the microstructure of the layers, as well as
without a need to employ any calculations based on distributed
fluctuating currents. Instead, the only physical characteristic on
which we base our theory is the effective input impedance of a stack
of layers, which can be obtained for any spatial harmonic of the field
(including both propagating and evanescent waves) using the methods of
VTLT. We have shown that such impedance representation, while being in
full agreement with sophisticated full-wave methods known from the
literature, results in simple formulas analogous to Nyquist
theory-based formulas for thermal noise in cascaded electric circuits
(for example, cascaded amplifiers). Therefore, with this model some
important concepts from the theory of electric networks
(conjugate-impedance match, optimal filtering, etc.) can be imported
into the field of radiative thermal transfer in multilayered
structures.

From the point of view of practical implementations, the developed
equivalent circuit approach offers significant simplifications as
compared to the known theories of radiative heat transfer based on
distributed fluctuating currents. Without any modifications, our
method can be used in heat transfer studies in uniaxial anisotropic
media that include micron and (or) submicron-thick layers. Moreover,
our model is applicable to radiative heat transfer in composite or
nanostructured layers (if these layers are effectively homogeneous for
spatial harmonics of the electromagnetic field which transfer the
radiative heat), and is readily generalizable to stratified
bi-anisotropic and spatially dispersive materials. Therefore, we hope
that our work may significantly enlarge the scope of the
radiative heat transfer research in composites, especially in nanostructured
metamaterials.  We believe that this may lead to new opportunities in
the design of efficient thermal energy harvesting devices, like
thermophotovoltaic converters and such.

\appendix

\section*{\label{App}Appendix: Nyquist formula for a reciprocal  anisotropic and lossy magneto-dielectric slab}

We consider a uniaxial magneto-dielectric slab described by the
macroscopic Maxwell equations for the time-harmonic fields \e
\l{Maxwell} \d\x\_E = -j\o\=\M_{\rm a}\.\_H - \_J^m, \quad \d\x\_H =
j\o\=\E_{\rm a}\.\_E + \_J^e,\f with the absolute permittivity and
permeability dyadics of the form $\=\E_{\rm a} =\E_0
(\E^{\perp}\=I_t+\E^{\parallel}\_z_0\_z_0)$ and $\=\M_{\rm a}
=\M_0(\M^{\perp}\=I_t+\M^{\parallel}\_z_0\_z_0)$. We assume that the
material of the slab is lossy, therefore, by the
fluctuation-dissipation theorem (at non-zero temperature) there appear
fluctuating external currents $\_J^e$ and $\_J^m$ in the slab. The
explicit form of these currents is not important at this stage.  As in
the main text, here we use the convention in which the time-harmonic
quantities are understood as root mean square (rms) values.

Let the fields $\_E'$, $\_H'$ be an arbitrary solution of the
Maxwell equations~\r{Maxwell} with $\_J^e = \_J^m = 0$ within the
slab. Then, considering the two systems of Maxwell equations with
non-zero sources and with vanishing sources, respectively, we can
form the Lorentz lemma \e \d\.(\_E\x\_H' - \_E'\x\_H) =
\_E'\.\_J^e - \_H'\.\_J^m.  \f Integrating it over the volume $V$
of the slab, we obtain the reciprocity relation \e \l{recipr}
\int_{S_1} \_n_1\.(\_E\x\_H' - \_E'\x\_H)\big|_{S_1}\,dS +
\int_{S_2} \_n_2\.(\_E\x\_H' - \_E'\x\_H)\big|_{S_2}\,dS =
\int_V(\_E'\.\_J^e - \_H'\.\_J^m)\,dV, \f where $S_{1,2}$ are at
the two interfaces of the slab, and $\_n_{1,2}$ are the outer unit
normals to these surfaces, respectively.

One may select any solution of the uniform Maxwell equations within
the slab for the fields $\_E'$, $\_H'$. For us it is convenient to use
the one that has the form \e \l{form}\_E'(\_r) =
\_E'_{-\_k_t}(z)e^{j\_k_t\.\_r}, \quad \_H'(\_r) =
\_H'_{-\_k_t}(z)e^{j\_k_t\.\_r}, \f where the $z$-axis is orthogonal
to the slab and the real vector $\_k_t$ lies in the plane of the slab
(the $xy$ plane). Physically, such a form corresponds to a
superposition of plane waves with the same transverse wavenumber:
$-\_k_t$. In Eq.~\r{form}, $\_E'_{-\_k_t}$ and $\_H'_{-\_k_t}$ define
the field solution profile within the slab as a function of $z$.

Substituting~\r{form} into the reciprocity relation~\r{recipr}, we
obtain (the first slab interface is at $z=z_1$ and the second one
is at $z=z_2$)
\begin{multline}
\l{recipr2} \_n_1\.(\_E_{\_k_t}\x\_H'_{-\_k_t} -
\_E'_{-\_k_t}\x\_H_{\_k_t})\big|_{z=z_1} +
\_n_2\.(\_E_{\_k_t}\x\_H'_{-\_k_t} -
\_E'_{-\_k_t}\x\_H_{\_k_t})\big|_{z=z_2} =\\
\int_{z_1}^{z_2}(\_E'_{-\_k_t}\.\_J^e_{\_k_t} -
\_H'_{-\_k_t}\.\_J^m_{\_k_t})\,dz,
\end{multline}
where we have decomposed the fluctuating currents $\_J^{e,m}$ and the
fields $\_E$, $\_H$ into plane waves using the Fourier transform
defined as \e \l{ksum} \_F(\_r) =
{A_0\over(2\pi)^2}\iint\_F_{\_k_t}(z)e^{-j\_k_t\.\_r}\,d^2\_k_t,\quad
\_F_{\_k_t}(z) = {1\over A_0}\iint\_F(\_r)e^{j\_k_t\.\_r}\,d^2\_r,\f
where $A_0$ is the unit area in the $xy$-plane, and $\_F$ can be any
of the fields or currents.

Eq.~\r{recipr2} is the reciprocity relation for the wave
components characterized with a fixed transverse wavenumber. In
order to simplify further writing we will use the notation
$\_F_{1,2} \equiv \_F_{\pm\_k_t}(z_{1,2})$ with $\_F$ being any of
the fields or currents. Then, noticing that only transverse
components of the fields play any role on the left-hand side
of~\r{recipr2} we rewrite it as
\begin{multline}
\l{lhs} \_n_1\.({\_E_1}_t\x{\_H'_1}_t - {\_E'_1}_t\x{\_H_1}_t) +
\_n_2\.({\_E_2}_t\x{\_H'_2}_t - {\_E'_2}_t\x{\_H_2}_t) =\\
{\_E'_1}_t\.(\_n_1\x{\_H_1}_t) - {\_E_1}_t\.(\_n_1\x{\_H'_1}_t) +
{\_E'_2}_t\.(\_n_2\x{\_H_2}_t) - {\_E_2}_t\.(\_n_2\x{\_H'_2}_t).
\end{multline}

Let us remind that the quantities ${\_E'_{1,2}}_t$ and
${\_H'_{1,2}}_t$ have the meaning of the transverse components of the
electric and magnetic fields at the interfaces of a source-free
magneto-dielectric slab. Therefore, as follows from the vector
transmission line theory (VTLT) for such slabs, these components are
related by the impedance matrix of the slab \e \l{zmatr}\vect{
  {\_E'_1}_t\\
  {\_E'_2}_t } = \matr{
  \=Z_{11} & \=Z_{12}\\
  \=Z_{21} & \=Z_{22} }\.\vect{
  \_n_1\x{\_H'_1}_t\\
  \_n_2\x{\_H'_2}_t }.\f The components of this matrix are dyadics
that are even in $\_k_t$: $\=Z_{mn}(-\_k_t) = \=Z_{mn}(\_k_t)$.  Also,
due to the symmetry and the reciprocity, $\=Z_{11}=\=Z_{22}$,
$\=Z_{12}=\=Z_{21}$, and $\=Z_{mn}^T=\=Z_{mn}$.~\cite{Tret}

Using~\r{zmatr} on the left-hand side of Eq.~\r{lhs} we obtain
\begin{multline}
\l{lhs2} A_0\Big[{\_E'_1}_t\.(\_n_1\x{\_H_1}_t) -
{\_E_1}_t\.(\_n_1\x{\_H'_1}_t) +
{\_E'_2}_t\.(\_n_2\x{\_H_2}_t) - {\_E_2}_t\.(\_n_2\x{\_H'_2}_t)\Big] =\\
\_I'_1\.\left[\=Z_{11}\.\_I_1+\=Z_{12}\.\_I_2 - \_V_1\right] +
\_I'_2\.\left[\=Z_{22}\.\_I_2+\=Z_{21}\.\_I_1 - \_V_2\right],
\end{multline}
where we have introduced the vector currents
$\_I'_{1,2}\equiv\sqrt{A_0}\,\_n_{1,2}\x{\_H'_{1,2}}_t$,
$\_I_{1,2}\equiv\sqrt{A_0}\,\_n_{1,2}\x{\_H_{1,2}}_t$ and the vector
voltages $\_V_{1,2}\equiv\sqrt{A_0}\,{\_E_{1,2}}_t$, and used the symmetry properties of
the impedance dyadics.

Let us now work on the right-hand side of~\r{recipr2}. At a fixed
$\_k_t$ the Maxwell equations for the fields $\_E'(\_r)$, $\_H'(\_r)$
reduce to a system of first-order linear differential equations for
the vector functions $\_E'_{-\_k_t}(z)$ and $\_H'_{-\_k_t}(z)$. From
the uniqueness theorem it follows that these functions are univocally
defined by boundary conditions imposed either on tangential electric
or tangential magnetic field.  Thus, we may consider two auxiliary
boundary-value problems, the first one with the boundary conditions \e
\l{bc1}\_n_1\x\_H^{\rm I}_{-\_k_t}(z_1) = \_I'_1/\sqrt{A_0}, \quad
\_n_2\x\_H^{\rm I}_{-\_k_t}(z_2) = 0, \f and the second one with \e
\l{bc2}\_n_1\x\_H^{\rm I}_{-\_k_t}(z_1) = 0, \quad \_n_2\x\_H^{\rm
  I}_{-\_k_t}(z_2) = \_I'_2/\sqrt{A_0}.  \f The field equations are
the same in these two problems.  From linearity it follows that the
superposition of the solutions of the two problems is the same as the
fields $\_E'_{-\_k_t}(z)$ and $\_H'_{-\_k_t}(z)$ that appear
in~\r{lhs2}. On the other hand, these problems physically correspond
to the two cases of the magneto-dielectric slab backed with a magnetic
wall (perfect magnetic conductor, PMC) at $z=z_2$ and at $z=z_1$,
respectively.

Let us consider the problem with the boundary conditions~\r{bc1}.
We may split the vector $\_I'_1$ into the components parallel and
orthogonal to $\_k_t$: \e \_I'_1 = I'_{1,\rm
TM}{\_k_t\over|\_k_t|} + I'_{1,\rm
TE}{\_k_t\x\_n_1\over|\_k_t|}.\f Thus, the component $I'_{1,\rm
TM}$ corresponds to TM-polarized field, and the component
$I'_{1,\rm TE}$ corresponds to TE-polarized field. Because the
wave equations in the slab also split into independent equations for
the TM and TE waves, we may also write for the vector fields \e
\_E^{\rm
  I}_{-\_k_t}(z) = \_E'_{1,\rm TM}(z) + \_E'_{1, \rm TE}(z), \quad
\_H^{\rm I}_{-\_k_t}(z) = \_H'_{1,\rm TM}(z) + \_H'_{1, \rm
TE}(z), \f where the addends are the TM and TE solutions for the
fields in the PMC-backed slab. The magnitudes of these solutions
are proportional to $I'_{1,\rm TM}$ and $I'_{1,\rm TE}$,
respectively.

Based on the above discussion we find for the right-hand side of~\r{recipr2}
\begin{multline}
\int_{z_1}^{z_2}(\_E^{\rm I}_{-\_k_t}\.\_J^e_{\_k_t} -
\_H^{\rm I}_{-\_k_t}\.\_J^m_{\_k_t})\,dz =\\
\int_{z_1}^{z_2}(\_E'_{1,\rm TM}\.\_J^e_{\_k_t} - \_H'_{1,\rm
TM}\.\_J^m_{\_k_t})\,dz + \int_{z_1}^{z_2}(\_E'_{1,\rm
TE}\.\_J^e_{\_k_t} -
\_H'_{1,\rm TE}\.\_J^m_{\_k_t})\,dz =\\
{1\over A_0}\,\_I'_1\.\left( e_{1,\rm TM}{\_k_t\over|\_k_t|} + e_{1,\rm TE}{\_k_t\x\_n_1\over|\_k_t|} \right),
\end{multline}
where \e \l{emf11} e_{1,\rm TM} = {A_0\over I'_{1,\rm
TM}}\int_{z_1}^{z_2}(\_E'_{1,\rm TM}\.\_J^e_{\_k_t} - \_H'_{1,\rm
TM}\.\_J^m_{\_k_t})\,dz, \f \e \l{emf12} e_{1,\rm TE} =
{A_0\over I'_{1,\rm TE}}\int_{z_1}^{z_2}(\_E'_{1,\rm
TE}\.\_J^e_{\_k_t} - \_H'_{1,\rm TE}\.\_J^m_{\_k_t})\,dz. \f In an
analogous manner we consider the second case with a
PMC at $z=z_1$ and find \e \int_{z_1}^{z_2}(\_E^{\rm
II}_{-\_k_t}\.\_J^e_{\_k_t} - \_H^{\rm
II}_{-\_k_t}\.\_J^m_{\_k_t})\,dz = {1\over A_0}\,\_I'_2\.\left( e_{2,\rm
TM}{\_k_t\over|\_k_t|} + e_{2,\rm
TE}{\_k_t\x\_n_2\over|\_k_t|} \right), \f where \e \l{emf21} e_{2,\rm TM} = {A_0\over I'_{2,\rm TM}}\int_{z_1}^{z_2}(\_E'_{2,\rm
TM}\.\_J^e_{\_k_t} - \_H'_{2,\rm TM}\.\_J^m_{\_k_t})\,dz, \f \e
\l{emf22} e_{2,\rm TE} = {A_0\over I'_{2,\rm
TE}}\int_{z_1}^{z_2}(\_E'_{2,\rm TE}\.\_J^e_{\_k_t} - \_H'_{2,\rm
TE}\.\_J^m_{\_k_t})\,dz. \f Therefore, combining these results
together and using~\r{recipr2}, \r{lhs}, and~\r{lhs2} we obtain
\begin{multline}
\l{emf1}
\_I'_1\.\left[\=Z_{11}\.\_I_1+\=Z_{12}\.\_I_2 - \_V_1\right] +
\_I'_2\.\left[\=Z_{22}\.\_I_2+\=Z_{21}\.\_I_1 - \_V_2\right] =\\
\_I'_1\.\left( e_{1,\rm TM}{\_k_t\over|\_k_t|} + e_{1,\rm TE}{\_k_t\x\_n_1\over|\_k_t|} \right) + \_I'_2\.\left(
e_{2,\rm TM}{\_k_t\over|\_k_t|} + e_{2,\rm
TE}{\_k_t\x\_n_2\over|\_k_t|} \right).
\end{multline}
Finally, because the vectors $\_I'_1$ and $\_I'_2$ are arbitrary, \e
\l{circm1} \matr{
  \=Z_{11} &\=Z_{12}\\
  \=Z_{21} &\=Z_{22}} \. \vect{\_I_1\\\_I_2} - \vect{\_V_1\\\_V_2} =
\vect{ \_e_1\\\_e_2}, \f where $\_e_{1,2}=e_{1,2,\rm
  TM}(\_k_t/|\_k_t|) + e_{1,2,\rm TE}{(\_k_t\x\_n_{1,2})/|\_k_t|}$.
These equations represent the equivalent vector circuit model of a
magneto-dielectric slab with fluctuating sources. In this model,
$\_I_{1,2}$ have the meaning of equivalent vector currents at the two
ports of a linear four-pole network of dyadic impedances, and
$\_e_{1,2}$ are the equivalent vector EMFs
acting at the two ports.

Because the equivalent EMFs are expressed through the fluctuating
currents, they are also fluctuating, stochastic quantities. As is
readily seen from~\r{emf11}--\r{emf12} and~\r{emf21}--\r{emf22}, the
stochastic mean value of the fluctuating EMFs is zero:
$\overline{\_e_{1,2}} = 0$, because $\overline{\_J^{e,m}_{\_k_t}} =
0$. However, the mean-square values of the fluctuating EMFs, as well
as their mutual correlations are in general different from zero and
can be calculated as follows:
\begin{multline}
\overline{(e_{\alpha,p}^*e_{\beta,q})} =\\
{A_0^2\over {I'}_{\alpha,p}^* {I'}_{\beta,q}}
\overline{
\int_{z_1}^{z_2}({\_E'}^*_{\alpha,p}\.{\_J^e_{\_k_t}}^* - {\_H'}^*_{\alpha,p}\.{\_J^m_{\_k_t}}^*)\,dz
\int_{z_1}^{z_2}({\_E'}_{\beta,q}\.{\_J^e_{\_k_t}} - {\_H'}_{\beta,q}\.{\_J^m_{\_k_t}})\,dz'
}=\\
{A_0^2\over {I'}_{\alpha,p}^* {I'}_{\beta,q}} \int_{z_1}^{z_2}\int_{z_1}^{z_2}
\overline{
({\_E'}^*_{\alpha,p}\.{\_J^e_{\_k_t}}^* - {\_H'}^*_{\alpha,p}\.{\_J^m_{\_k_t}}^*)\big|_z\,
({\_E'}_{\beta,q}\.{\_J^e_{\_k_t}} - {\_H'}_{\beta,q}\.{\_J^m_{\_k_t}})\big|_{z'}
}
\,dz\,dz'=\\
{A_0^2\over {I'}_{\alpha,p}^* {I'}_{\beta,q}}\left[
\int_{z_1}^{z_2}\int_{z_1}^{z_2}
{\_E'}^*_{\alpha,p}(z)\.\overline{
\_J^{e\,*}_{\_k_t}(z)\_J^{e}_{\_k_t}(z')
}
\.{\_E'}_{\beta,q}(z')
\,dz\,dz'+\right.\\
\left.\int_{z_1}^{z_2}\int_{z_1}^{z_2}
{\_H'}^*_{\alpha,p}(z)\.
\overline{
\_J^{m\,*}_{\_k_t}(z)\_J^{m}_{\_k_t}(z')
}
\.\_H'_{\beta,q}(z')
\,dz\,dz'\right],
\l{msemf}
\end{multline}
where $\alpha,\beta = 1,2$ and $ p,q = \rm TE,TM$. There are no cross
terms in the last integral of Eq.~\r{msemf} because the electric and
magnetic fluctuations are statistically independent: the dyadic
$\_J^{e\,*}_{\_k_t}{\_J^m_{\_k_t}}$ is such that
$\overline{\_J^{e\,*}_{\_k_t}{\_J^m_{\_k_t}}} = 0$.

From the fluctuation-dissipation theorem, for the fluctuating
currents composed of harmonics within a narrow interval around a
given frequency $\o$,
\begin{align}
\l{fdt1} \overline{\_J^{e\,*}_{\_k_t}(z)\_J^{e}_{\_k_t}(z')} &=
{1\over\pi A_0}j\o\left(\=\E_{\rm a}-\=\E_{\rm a}^\dagger\right)
\delta(z-z')\Theta(\o,T)\Delta\o,\\
\l{fdt2} \overline{\_J^{m\,*}_{\_k_t}(z)\_J^{m}_{\_k_t}(z')} &=
{1\over\pi A_0}j\o\left(\=\M_{\rm a}-\=\M_{\rm a}^\dagger\right)
\delta(z-z')\Theta(\o,T)\Delta\o.
\end{align}
The dimensionality factor $1/A_0$ appears in~\r{fdt1}--\r{fdt2}
because of the form of transformation~\r{ksum}. Note also
that~\r{fdt1}--\r{fdt2} are written for the rms amplitudes of the
fluctuating currents.

Substituting~\r{fdt1}--\r{fdt2} into \r{msemf}
and evaluating the integrals over $z'$ we find that \e
\overline{(e_{\alpha,p}^*e_{\beta,q})} = {j\o
  A_0\Theta(\o,T)\Delta\o\over\pi {I'}^*_{\alpha,p} {I'}_{\beta,q}}
\int_{z_1}^{z_2}\left[
  {\_E'}^*_{\alpha,p}\.\left(\=\E_{\rm a}-\=\E_{\rm a}^\dagger\right)\.{\_E'}_{\beta,q}
  +
  {\_H'}^*_{\alpha,p}\.\left(\=\M_{\rm a}-\=\M_{\rm a}^\dagger\right)\.{\_H'}_{\beta,q}
\right]\,dz. \l{corrint}\f However, from the well-known differential
lemma \e \d\.(\_E_1\x\_H_2^* + \_E_2^*\x\_H_1) =
-j\o\left[\_E_2^*\.\left(\=\E_{\rm a}-\=\E_{\rm a}^\dagger\right)\.\_E_1 +
  \_H_2^*\.\left(\=\M_{\rm a}-\=\M_{\rm a}^\dagger\right)\.\_H_1\right], \f which
holds for arbitrary source-free electromagnetic fields $\_E_{1,2}(\_r)$,
$\_H_{1,2}(\_r)$ within the slab, it follows that
\begin{multline}
\l{mutualpower}
\int_{z_1}^{z_2}\left[
  {\_E'}^*_{\alpha,p}\.\left(\=\E_{\rm a}-\=\E_{\rm a}^\dagger\right)\.{\_E'}_{\beta,q}
  +
  {\_H'}^*_{\alpha,p}\.\left(\=\M_{\rm a}-\=\M_{\rm a}^\dagger\right)\.{\_H'}_{\beta,q}
\right]\,dz =\\
-{1\over j\o}\left[
\_n_1\.({\_E'}_{\beta,q}\x{\_H'}^*_{\alpha,p})\big|_{z=z_1} +
\_n_1\.({\_E'}^*_{\alpha,p}\x{\_H'}_{\beta,q})\big|_{z=z_1} +\right.\\
\left.\_n_2\.({\_E'}_{\beta,q}\x{\_H'}^*_{\alpha,p})\big|_{z=z_2} +
\_n_2\.({\_E'}^*_{\alpha,p}\x{\_H'}_{\beta,q})\big|_{z=z_2}
\right],
\end{multline}
from which we see that if $p \neq q$, the integral~\r{mutualpower}
vanishes due to the orthogonality of the TE and TM
polarizations. Next, when $p=q$ and $\alpha = \beta = 1$ we obtain
from \r{mutualpower}, \r{bc1}--\r{bc2}, and \r{zmatr}: \e
\int_{z_1}^{z_2}\left[
  {\_E'}^*_{1,p}\.\left(\=\E_{\rm a}-\=\E_{\rm a}^\dagger\right)\.{\_E'}_{1,p} +
  {\_H'}^*_{1,p}\.\left(\=\M_{\rm a}-\=\M_{\rm a}^\dagger\right)\.{\_H'}_{1,p}
\right]\,dz ={2|I'_{1,p}|^2\over j\o A_0}\Re(Z_{11}^p).\f An analogous
result is obtained for $\alpha = \beta = 2$. On the other hand, when
$\alpha=1$ and $\beta=2$ we obtain \e \int_{z_1}^{z_2}\left[
  {\_E'}^*_{1,p}\.\left(\=\E_{\rm a}-\=\E_{\rm a}^\dagger\right)\.{\_E'}_{2,p} +
  {\_H'}^*_{1,p}\.\left(\=\M_{\rm a}-\=\M_{\rm a}^\dagger\right)\.{\_H'}_{2,p}
\right]\,dz ={{I'}_{1,p}^*I'_{2,p}\over j\o A_0}\left(Z_{12}^p +
  {Z_{21}^p}^*\right).  \f Combining all these results together and
using the reciprocity property of the $Z$-parameters we find
from~\r{corrint} that \e\l{genNyquist}
\overline{(e_{\alpha,p}^*e_{\beta,p})}=
{2\over\pi}\Re\left(Z_{\alpha\beta}^p\right)\Theta(\o,T)\Delta\o,\f
where $\alpha,\beta = 1,2$ and $p = \rm TE, TM$. Eq.~\r{genNyquist} is
the generalized Nyquist formula for the thermal-electromagnetic noise
in a uniaxial magneto-dielectric layer.

\end{document}